\documentclass[twocolumn,journal, bookmarks=false]{IEEEtran}
\IEEEoverridecommandlockouts

\usepackage{color} 
\usepackage{graphicx}
\usepackage{amsmath}
\usepackage{amssymb}
\usepackage{algorithm}
\usepackage{algorithmic}
\usepackage{amsmath}
\usepackage{multirow}
\usepackage{booktabs}
\usepackage{array}
\usepackage{amsthm}
\usepackage{stfloats}
\usepackage{caption}
\usepackage{subfigure}
\usepackage{bm}
\usepackage{epstopdf}
\usepackage{setspace}
\usepackage{gensymb}
\usepackage{url}
\usepackage{verbatim}
\usepackage{balance}
\usepackage{xcolor}

\usepackage{hyperref}
\usepackage{cleveref}

\hypersetup{colorlinks = true,
    linkcolor = black,
    urlcolor = black,
    citecolor = black}

\definecolor{color_radar}{rgb}{0.298, 0.596, 0.902} 
\definecolor{color_comm}{rgb}{0.690, 0.863, 0.400} 
\newtheorem{prop}{Proposition}

\newcommand{\be}{\begin{equation}}
        \newcommand{\ee}{\end{equation}}
\newcommand{\bea}{\begin{eqnarray}}
        \newcommand{\eea}{\end{eqnarray}}
\newcommand{\ba}{\begin{array}}
        \newcommand{\ea}{\end{array}}

\allowdisplaybreaks[4]

\title{Sensing-Oriented Adaptive Resource Allocation Designs for OFDM-ISAC Systems
    \thanks{P. Li and M. Li are with the School of Information and Communication Engineering, Dalian University of Technology, Dalian 116024, China (e-mail: lipeishi@mail.dlut.edu.cn; mli@dlut.edu.cn).}
    \thanks{R. Liu and A. Lee Swindlehurst are with the Center for Pervasive Communications and Computing, University of California, Irvine, CA 92697, USA (e-mail: rangl2@uci.edu; swindle@uci.edu).}
    \thanks{Q. Liu is with the School of Computer Science and Technology, Dalian University of Technology, Dalian 116024, China (e-mail: qianliu@dlut.edu.cn).}
}
\author{Peishi Li,~\IEEEmembership{Graduate Student Member,~IEEE,}
    Ming Li,~\IEEEmembership{Senior Member,~IEEE,}
    Rang Liu,~\IEEEmembership{Member,~IEEE,}  
    Qian Liu,~\IEEEmembership{Member,~IEEE,}
    and A. Lee Swindlehurst, ~\IEEEmembership{Fellow,~IEEE}}

\begin{document}

\maketitle
\thispagestyle{empty}
\begin{abstract}
Orthogonal frequency division multiplexing - integrated sensing and communication (OFDM-ISAC) has emerged as a key enabler for future wireless networks, leveraging the widely adopted OFDM waveform to seamlessly integrate wireless communication and radar sensing within a unified framework. In this paper, we propose adaptive resource allocation strategies for OFDM-ISAC systems to achieve optimal trade-offs between diverse sensing requirements and communication quality-of-service (QoS). We first develop a comprehensive resource allocation framework for OFDM-ISAC systems, deriving closed-form expressions for key sensing performance metrics, including delay resolution, Doppler resolution, delay-Doppler peak sidelobe level (PSL), and received signal-to-noise ratio (SNR). Building on this theoretical foundation, we introduce two novel resource allocation algorithms tailored to distinct sensing objectives. The resolution-oriented algorithm aims to maximize the weighted delay-Doppler resolution while satisfying constraints on PSL, sensing SNR, communication sum-rate, and transmit power. The sidelobe-oriented algorithm focuses on minimizing delay-Doppler PSL while satisfying resolution, SNR, and communication constraints. To efficiently solve the resulting non-convex optimization problems, we develop two adaptive resource allocation algorithms based on Dinkelbach's transform and majorization-minimization (MM). Extensive simulations validate the effectiveness of the proposed sensing-oriented adaptive resource allocation strategies in enhancing resolution and sidelobe suppression. Remarkably, these strategies achieve sensing performance nearly identical to that of a radar-only scheme, which dedicates all resources to sensing. These results highlight the superior performance of the proposed methods in optimizing the trade-off between sensing and communication objectives within OFDM-ISAC systems.

\end{abstract}

\vspace{1mm}

\section{Introduction}
The proliferation of applications such as autonomous driving, smart cities, and low-altitude economics has driven significant advances in wireless networks, demanding high-speed connectivity and precise environmental awareness. Integrated sensing and communication (ISAC) has emerged as a transformative paradigm to address these dual requirements by enabling efficient sharing of hardware and spectrum between communication and sensing functionalities \cite{LiuFan_JSAC_2022}-\cite{Zhang_JSTSP_2021}. By harmonizing multi-domain resource allocation, ISAC achieves significant reductions in infrastructure cost and improvements in spectral efficiency. These advancements position ISAC as a foundational enabler for sixth-generation (6G) networks  \cite{Zhang_CST_2022}, \cite{LiuRang JSTSP 2022}.

Orthogonal frequency division multiplexing (OFDM) has been widely adopted in wireless communication standards due to its high spectral efficiency and resilience to frequency-selective fading. Beyond its advantages for communication, OFDM also exhibits strong potential for radar sensing, as its inherent frequency diversity enables accurate target detection and parameter estimation \cite{Sturm_Proc_2011}, \cite{Hakobyan_SPM_2019}. Leveraging this dual capability, OFDM serves as a promising waveform for ISAC in future 6G networks, facilitating seamless integration of sensing and communication of both functionalities without necessitating major modifications to existing protocols or hardware architectures.

However, in contrast to deterministic radar probing signals designed specifically for optimal sensing performance, conventional OFDM communication waveforms exhibit randomness due to the inherently random data embedded within them. This randomness can significantly degrade sensing performance by introducing undesirable fluctuations in the ambiguity function (AF), thereby impairing the accuracy of target detection and parameter estimation \cite{Keskin_TSP_2021}. As a result, a fundamental trade-off arises between the benefit of using deterministic sensing signals versus the need to incorporate randomness in the waveform to transmit information. This poses a critical challenge to the design of OFDM-ISAC systems.

To address this conflict, a natural approach is to reshape the random communication waveform by optimizing its AF with respect to its statistical characteristics \cite{Du_TSP_2024}, \cite{Fan_Liu_2025} or its specific instantaneous properties \cite{Peishi_TWC_2025}, \cite{Peishi_TVT_2025}, in order to facilitate its use in sensing. While these methods offer a trade-off between communication randomness and sensing determinism, they typically require modifications to conventional OFDM modulation schemes and involve complex waveform designs, posing significant challenges for practical OFDM-ISAC implementation. In contrast, a more practical approach exploits the flexible resource allocation capabilities of the OFDM framework, enabling communication and sensing signals to occupy distinct time-frequency resources. This strategy effectively mitigates the detrimental impact of random communication symbols on sensing performance while ensuring spectral coexistence. Thanks to its straightforward implementation and strong compatibility with existing OFDM-based communication systems, this approach has garnered significant attention in the OFDM-ISAC domain.

Resource allocation strategies have been extensively explored in radar systems in order to enhance specific sensing performance requirements under resource constraints \cite{Sun_RadConf_2014}-\cite{Hakobyan_TAES_2020}. Building on these foundations, recent studies have leveraged the flexibility of OFDM systems to allocate resources for both communication and sensing functions. In particular, several subcarrier selection and power allocation algorithms have been proposed that jointly optimize radar and communication performance  \cite{Bica_ICASSP_2019}-\cite{Chen_CL_2023}. However, these approaches predominantly focus on frequency-domain resource allocation for range estimation and neglect time-domain optimization, which is crucial for velocity (Doppler) estimation performance. Recent work \cite{ZhangFan_JSAC_2023} has considered resource allocation optimization across both domains, although this work emphasizes sidelobe suppression to enhance sensing performance. While sidelobe suppression is important for mitigating weak target masking and ambiguities, it alone is insufficient to achieve satisfactory sensing performance for different applications. The sensing resolution, which is typically determined by the width of the main lobe of the AF, is another crucial metric for distinguishing closely spaced targets. Overlooking the interdependencies among various AF features can lead to suboptimal trade-offs between resolution, sidelobe suppression, and ambiguity mitigation. Moreover, sensing performance depends not only on the AF characteristics of the transmitted waveform, but also on the quality of the received echo signal, which is subject to noise. Consequently, the sensing signal-to-noise ratio (SNR) also plays a pivotal role in enhancing detection accuracy and estimation precision, and must be incorporated into the resource allocation process.

To address these challenges, this paper proposes novel adaptive resource allocation designs for OFDM-ISAC systems, considering multiple sensing performance metrics. The key contributions are summarized as follows:
\begin{itemize}  
\item We establish a comprehensive resource allocation framework for OFDM-ISAC systems, deriving closed-form expressions for several key sensing performance metrics, including delay resolution, Doppler resolution, delay-Doppler PSL, and received signal-to-noise ratio (SNR). These metrics provide a comprehensive characterization of sensing performance, forming the theoretical basis for optimizing resource allocation in OFDM-ISAC systems.

\item For sensing scenarios where resolution is the primary concern, we propose a resolution-oriented resource allocation algorithm. This algorithm maximizes the weighted delay-Doppler resolution while ensuring that constraints on PSL, sensing SNR, communication sum-rate, and transmit power budget are satisfied.

\item For sensing applications where sidelobe interference that masks weaker targets is the primary concern, we introduce a sidelobe-oriented resource allocation algorithm. This algorithm focuses on minimizing PSL while satisfying constraints on delay and Doppler resolutions, sensing SNR, communication sum-rate, and transmit power budget. To efficiently solve these non-convex problems, we develop two optimization algorithms based on Dinkelbach's transform and the majorization-minimization (MM) method.

\item Extensive simulations demonstrate the effectiveness of the proposed adaptive resource allocation strategies in comparison to conventional resource allocation methods. For example, the resolution-oriented algorithm achieves a $5$dB SNR improvement in resolution at the same root-mean-square-error (RMSE) level for range and velocity estimation. Meanwhile, the sidelobe-oriented algorithm results in a $90$dB reduction in PSL, significantly improving range and velocity estimation performance, particularly in multi-target sensing scenarios. Notably, the proposed adaptive resource allocation designs achieve nearly the same sensing performance as the radar-only scheme, which dedicates all resources to sensing. These results highlight the ability of the proposed methods to strike a balanced trade-off between sensing and communication performance.
\end{itemize}

\textit{Notation:} Unless otherwise specified, the following notation is used throughout the paper. Boldface lower-case letters (e.g., $\mathbf{x}$) indicate column vectors, while bold upper-case letters (e.g., $\mathbf{X}$) indicate matrices. The sets $\mathbb{C}$ and $\mathbb{Z}$ represent the collection of complex numbers and integers, respectively. Superscripts $()^{\ast}$, $()^{T}$, and $()^{H}$ indicate the conjugate, transpose, and transpose-conjugate, respectively. The operators $\mathfrak{R}\{ \cdot \}$ and $\mathfrak{I}\{ \cdot \}$ extract the real and imaginary parts of a complex number. An $N\times N$ identity matrix is denoted by $\mathbf{I}_{N}$, while $\mathbf{1}_{N}$ and $\mathbf{0}_{N}$ are $N \times 1$ vectors with all-one or all-zero entries, respectively. The $\ell_{2}$ norm of a vector is indicated by $\| \cdot \|$. The function $\text{vec} \{ \mathbf{X} \}$ vectorizes the matrix $\mathbf{X}$ column-by-column. The operators $\otimes$ and $\odot$ represent the Kronecker product and Hadamard (element-wise) product, respectively. The notation $\mathbf{A}(i, j)$ indicates the $(i, j)$-th entry of matrix $\mathbf{A}$. 

\section{System Model}
\subsection{Transmit Signal Model}
We consider a monostatic OFDM-ISAC system, where a dual-function base station employs OFDM signals to serve $K$ single-antenna communication users while simultaneously performing radar sensing using the target echo signals. The OFDM frame consists of $N$ subcarriers and $M$ symbols, and the $n$-th subcarrier of the $m$-th symbol is defined as the $(n, m)$-th resource element (RE). To avoid the adverse effects of random communication symbols on sensing performance, the OFDM time-frequency resources are partitioned into two mutually exclusive sets: one dedicated to radar sensing function and the other to communication purposes.
Specifically, we employ $\mathbf{U}_0 \in \{0, 1\}^{N \times M}$ and $\mathbf{U}_{k} \in \{0, 1\}^{N \times M}$ as the RE selection indicators for radar sensing and the $k$-th communication user, $k = 1,2, \dots, K$,  respectively. The $(n,m)$-th entries of $\mathbf{U}_0$ and $\mathbf{U}_{k}$ are respectively defined as
\begin{equation}
    \begin{aligned}
    & u_{n,m,0} \triangleq
        \begin{cases}
            1 & ~\text{if the $(n,m)$-th RE is for sensing}; \\
            0 & ~\text{otherwise},
        \end{cases} \\
    & u_{n,m,k} \triangleq
        \begin{cases}
            1 & ~\text{if the $(n,m)$-th RE is for the $k$-th user}; \\
            0 & ~\text{otherwise}.
        \end{cases}
    \end{aligned}
\end{equation}
To avoid mutual interference and ensure the exclusivity of the resource allocation, each RE can only be occupied by either radar sensing or a single communication user, which means
\begin{equation}
    \mathbf{U}_0 + \sum_{k=1}^{K} \mathbf{U}_k \preceq  \mathbf{1}_{N \times M}.
\end{equation}

For communication, the information data matrix is denoted as $\mathbf{S}_{\text{c}}$ with $s_{n,m}^{\text{c}}$ representing the modulated symbol on the $(n, m)$-th RE, and each symbol is drawn from a set of fixed constellations. For sensing, the dedicated radar probing signal matrix is denoted as $\mathbf{S}_{\text{r}}$. In this paper, we employ classical Zadoff-Chu sequences as the dedicated radar probing signal owing to its ideal periodic auto-correlation property \cite{Chu_TIT_1972}. In particular, the $(n,m)$-th entry of $\mathbf{S}_{\text{r}}$ is given by
\begin{equation}
    s_{n,m}^{\text{r}} = e^{- \jmath \pi q_m \frac{n(n+1)}{N}}, ~\forall n \in \mathcal{N}, m \in \mathcal{M},
\end{equation}
where $q_m$ is the root index of the Zadoff-Chu sequence, and where $\mathcal{N} = \{0, 1, \dots, N-1\}$ and $\mathcal{M} = \{0, 1, \dots, M-1\}$ denote the index sets of the OFDM subcarriers and symbols, respectively.

Let $p_{n,m}$ represent the transmit power on the $(n,m)$-th RE. Then, the compound OFDM transmit signal matrix $\mathbf{X} \in \mathbb{C}^{N \times M}$ can be written as
\begin{equation}
    \mathbf{X} = \mathbf{A} \odot \Big( \mathbf{U}_0 \odot \mathbf{S}_{\text{r}} + \sum_{k=1}^K \mathbf{U}_k \odot \mathbf{S}_{\text{c}} \Big),
\end{equation}
where $\mathbf{A}(n,m) = \sqrt{p_{n,m}}$. The modulated baseband OFDM time-domain transmit signal can be expressed as
\begin{equation}
    x(t) = \frac{1}{\sqrt{N}}  \sum_{m = 0}^{M-1} \sum_{n = 0}^{N-1} x_{n,m} e^{\jmath 2 \pi n \Delta_f t} g\bigg( \frac{t_m}{T_{\text{sym}}} \bigg),
\end{equation}
where $\Delta_{f} = 1/T$ denotes the subcarrier spacing, $T$ denotes the OFDM symbol duration, $T_{\text{sym}} = T + T_{\text{CP}}$ is the total symbol duration including the cyclic prefix (CP) of length $T_{\text{CP}}$, $t_m=t - m T_{\text{sym}}$ is the relative fast-time variable during the $m$-th OFDM symbol, and $g(\cdot)$ denotes the rectangular pulse that equals $1$ for $t \in [0, 1]$ and $0$ otherwise. Before transmission, the baseband OFDM signal is up-converted to the radio frequency domain, 
\begin{equation}
    \widetilde{x}(t) = \mathfrak{R}\big\{ x(t) e^{\jmath 2 \pi f_{\text{c}} t}\big\}  ,
\end{equation}
where $f_{\text{c}}$ denotes the carrier frequency.

\subsection{Received Signal Model for Multi-user Communications}
The received signal at each user is down-converted to baseband, followed by analog-to-digital conversion, CP removal, and a discrete Fourier transform (DFT) operation. Finally, the received communication signal on the $(n,m)$-th RE of the $k$-th user can be written as
\begin{equation}
    y_{n,m,k} =  h_{n,m,k}\sqrt{p_{n,m}}u_{n,m,k} s_{n,m}^{\text{c}} + w_{n,m,k},
\end{equation}
where $h_{n,m,k}$ is the frequency domain communication channel, $w_{n,m,k} \sim \mathcal{CN} (0, \sigma^2)$ denotes additive white Gaussian noise (AWGN), and $\sigma^2 = N_0 \Delta_f$ with $N_0$ representing the noise power spectral density (PSD). The received signal at the $k$-th user $\mathbf{Y}_k$ with $\mathbf{Y}_k(n,m) = y_{n,m,k}$ can be expressed as
\begin{equation}
    \mathbf{Y}_k = \mathbf{H}_k \odot \mathbf{A} \odot \mathbf{U}_k \odot \mathbf{S}_{\text{c}} + \mathbf{W}_k ,
\end{equation}
where $\mathbf{H}_k \in \mathbb{C}^{N \times M}$ with $\mathbf{H}_k(n,m) = h_{n,m,k}$ denotes the communication channel of the $k$-th user, and $\mathbf{W}_k(n,m) = w_{n,m,k}$. The average achievable sum-rate of the multi-user communication system can thus be formulated as
\begin{equation}
    \!\!\! R_{\text{c}} = \frac{1}{MN} \! \sum_{k=1}^{K} \! \sum_{m=0}^{M-1} \! \sum_{n=0}^{N-1} \! \log_2 \bigg(1 + \frac{ |h_{n,m,k}|^2p_{n,m}u_{n,m,k} }{\sigma^2} \bigg).
\end{equation}

\subsection{Received Signal Model for Radar Sensing}
Suppose there exist $Q$ point targets at ranges $R_q$ with relative radial velocities $v_q$ for $q = 1, \dots, Q$. Then the echo signal at the sensing receiver can be expressed as
\begin{equation}
    y(t) = \sum_{q=1}^{Q} \alpha_q x(t - \tau_q)e^{\jmath 2 \pi f_{\text{d}, q} t} + z(t),
\end{equation}
where $\alpha_q$, $\tau_q = 2 R_q / c_0$, $f_{\text{d}, q} = 2v_q / \lambda$ denote the complex channel gain including the path loss and radar cross section (RCS), round-trip delay, and Doppler shift of the $q$-th target, respectively. Here, $\lambda = c_0/f_{\text{c}}$ is the wavelength, $c_0$ is the speed of light, and $z(t) \sim \mathcal{CN}(0, \sigma^2)$ denotes AWGN. The channel gain is derived from the radar range equation $|\alpha_q|^2 = \frac{\sigma_{\text{rcs}, q} \lambda^2}{ (4\pi)^3 R_q^4 }$, where $\sigma_{\text{rcs}, q}$ denotes the RCS of the $q$-th target, which is modeled as $\sigma_{\text{rcs}, q} \sim \text{exp}(\varsigma)$ under the Swerling I model \cite{Swerling_TAES_1997}. After the CP removal and DFT operation, the frequency-domain echo signal corresponding to the $(n,m)$-th RE is given by \cite{Braun_thesis_2014}
\begin{equation} \label{eq:echo_freq_sig}
    y_{n,m} = \sum_{q=1}^{Q} \alpha_q e^{-\jmath 2 \pi n \Delta_f \tau_q} e^{\jmath 2 \pi m f_{\text{d}, q} T_{\text{sym}}} x_{n,m} + z_{n,m},
\end{equation}
where $z_{n,m}$ is noise. For clarity, we respectively reformulate the sensing channel and the radar echo signal in matrix notation as
\begin{subequations}
    \begin{align}
        \mathbf{H}_0 & = \sum_{q=1}^Q \alpha_q \boldsymbol{\phi}(\tau_q) \boldsymbol{\psi}^H(f_{\text{d}, q}), \\
        \mathbf{Y}_0 & = \mathbf{H}_0 \odot \mathbf{X}  + \mathbf{Z},
    \end{align}
\end{subequations}
where $\mathbf{Y}_0 \in \mathbb{C}^{N \times M}$ with $\mathbf{Y}_0(n,m) = y_{n,m}$, $\mathbf{Z} \in \mathbb{C}^{N \times M}$ with $\mathbf{Z}(n,m) = z_{n,m}$, and $\boldsymbol{\phi}(\tau_q)$ and $\boldsymbol{\psi}(f_{\text{d}, q})$ denote the steering vectors in the delay and Doppler domains, defined as
\begin{subequations}
    \begin{align}
    \boldsymbol{\phi}(\tau_q)  & \triangleq [1, e^{-\jmath 2 \pi \Delta_f \tau_q}, \dots, e^{-\jmath 2 \pi (N-1) \Delta_f \tau_q}]^T,    \\
    \boldsymbol{\psi}(f_{\text{d}, q}) & \triangleq [1, e^{-\jmath 2 \pi f_{\text{d}, q} T_{\text{sym}}}, \dots, e^{-\jmath 2 \pi (M-1) f_{\text{d}, q} T_{\text{sym}}}]^T.
    \end{align}
\end{subequations}

To mitigate the adverse effects of the randomness in the communication data $\mathbf{S}_{\text{c}}$ on the radar AF, we exclusively use the sensing REs for target detection and estimation. The observations on the sensing REs can be expressed as
\begin{equation} \label{eq:echo_sig_sensingRE}
    \begin{aligned}
        \hspace{-2mm}\mathbf{Y}_{\text{r}}
         & = \mathbf{U}_0 \odot \mathbf{Y}_0  \\
         & = \mathbf{U}_0 \odot (\mathbf{H}_0 \odot \mathbf{A} \odot \mathbf{S}_{\text{r}} + \mathbf{Z}).
    \end{aligned}
\end{equation}
Based $\mathbf{Y}_{\text{r}}$, the radar can perform a two-dimensional DFT to obtain the range-velocity image $\boldsymbol{\chi}^{\text{RD}}$, i.e.
\begin{equation}
    \boldsymbol{\chi}^{\text{RD}} = \mathbf{F}_N^H (\mathbf{Y}_{\text{r}} \odot \mathbf{S}_{\text{r}}^{\ast}) \mathbf{F}_M,
\end{equation}
where $\mathbf{F}_N^H$ and $\mathbf{F}_M$ denote the normalized DFT and inverse-DFT (IDFT) matrix operators, respectively. Then, range-velocity parameter estimation can be performed by searching for peaks in the range-velocity image, e.g., using a constant false alarm rate (CFAR) detector. In the next section, we will analyze the impact of RE selection and power allocation strategies on sensing performance and derive the corresponding sensing performance metrics.

\begin{figure*}[!t]
    \centering
    \includegraphics[width = 7.2 in]{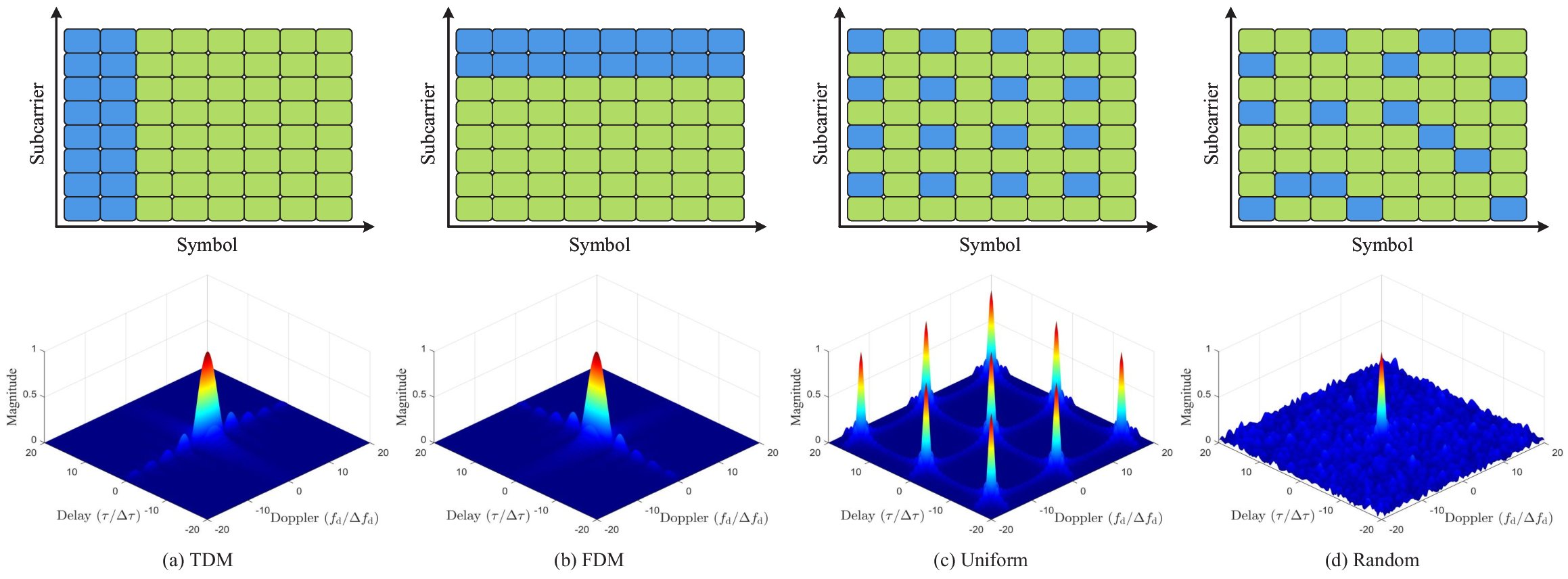}
    \caption{Common time-frequency resource allocation schemes in OFDM-ISAC systems (sensing: \textcolor{color_radar}{\rule{7pt}{7pt}}, communication: \textcolor{color_comm}{\rule{7pt}{7pt}}), along with their corresponding AFs.}\label{fig:AF_benchmarks}
    \vspace{-2mm}
\end{figure*}

\section{Sensing Performance Metrics}
The AF is a fundamental concept in radar signal processing, used to characterize a radar system's ability to distinguish between targets at different ranges and velocities. In the following, we first conduct a detailed analysis of the AF for the considered OFDM-ISAC system, and then derive radar sensing metrics, including resolution, sidelobe level, as well as the sensing receive signal-to-noise ratio (SNR).  

\subsection{AF Analysis}
The AF is defined as the time-frequency composite correlation of the transmit signal. Since the sensing receiver only utilizes the echo signal from the sensing REs, we focus on the AF of the dedicated radar probing signal, which can be expressed as
\begin{equation} \label{eq:radar_tx_sig}
    x_{\text{r}}(t) = \frac{1}{\sqrt{N}}  \sum_{m = 0}^{M-1} \sum_{n = 0}^{N-1} u_{n,m}^0 s_{n,m}^{\text{r}} e^{\jmath 2 \pi n \Delta_f t} g\bigg( \frac{t_m}{T_{\text{sym}}} \bigg).
\end{equation}
Using the definition of the AF in \cite{woodward_book_1953}:
\begin{equation} \label{eq:AF_define}
    \chi(\tau,f_{\text{d}}) \triangleq \int_{-\infty}^{\infty} x_{\text{r}}(t)  x_{\text{r}}^{\ast}(t+\tau)e^{\jmath 2 \pi f_{\text{d}}t} \text{d}t,
\end{equation}
and by substituting (\ref{eq:radar_tx_sig}) into (\ref{eq:AF_define}), we obtain a compact expression for the AF, which is stated in the following proposition.
\begin{prop} \label{prop:AF_deri}
    The AF of the OFDM signal with selected sensing REs can be written as
    \begin{equation} \label{eq:AF_expression}
        \chi(\tau, f_{\text{d}}) = \frac{1}{MN} \sum_{m=0}^{M-1} \sum_{n=0}^{N-1} u_{n, m}^0  p_{n,m} e^{-\jmath 2 \pi n \Delta_f \tau} e^{\jmath 2 \pi m f_{\text{d}} T_{\text{sym}}}.
    \end{equation}
\end{prop}
\noindent \textit{Proof:} See Appendix \ref{appendix:AF_deri}.
\hfill $\blacksquare$
\vspace{\baselineskip}

For the convenience in the following analysis, we define
\begin{equation}
    \begin{aligned}
    \mathbf{u}_0  & \triangleq \text{vec}(\mathbf{U}_0) \in \{0, 1\}^{MN},  \hspace{1mm} \mathbf{p} \! \triangleq \! [p_{0,0}, p_{1, 0}, \dots, p_{N\!-\!1, M\!-\!1}]^T,   \\
    \boldsymbol{\Phi}_{\tau} & \triangleq \text{diag}\{\boldsymbol{\mathbf{\phi}}(\tau)\} \in \mathbb{C}^{N \times N}, \hspace{1mm} \boldsymbol{\Psi}_{f_{\text{d}}} \triangleq \text{diag}\{\boldsymbol{\mathbf{\psi}}(f_{\text{d}})\} \in \mathbb{C}^{M \times M}.
    \end{aligned}
\end{equation}
Accordingly, the AF in (\ref{eq:AF_expression}) can be rewritten in a compact form as
\begin{equation}
    \chi(\tau, f_{\text{d}}) = \frac{1}{MN} \mathbf{u}_0^T (\boldsymbol{\Psi}_{f_{\text{d}}}^H \otimes \boldsymbol{\Phi}_{\tau}) \mathbf{p}.
\end{equation}

The distribution of the sensing REs significantly impacts the resolution and sidelobe level of the AF, ultimately influencing the parameter estimation capability. To illustrate this fact, Fig. \ref{fig:AF_benchmarks} presents four commonly used sensing RE allocation schemes in OFDM-ISAC systems: time division multiplexing (TDM), frequency division multiplexing (FDM), uniform allocation, and random allocation, along with the corresponding AFs. We summarize their characteristics and sensing performance as follows:
\begin{itemize}
    \item \textbf{TDM}: Here, the sensing REs occupy a contiguous set of OFDM symbols. This allocation leads to poor Doppler resolution due to the limited use of slow-time domain resources.
    \item \textbf{FDM}: Similar to TDM, FDM 
    assigns a contiguous block of OFDM subcarriers to the sensing REs. However, insufficient use of frequency domain resources leads to degraded delay resolution.
    \item \textbf{Uniform}: In this approach, the sensing REs are uniformly distributed across the time-frequency domain. While this scheme mitigates the limitations of TDM and FDM, insufficient sampling introduces periodic shifts in the delay and Doppler spectra, reducing the maximum unambiguous range and velocity.
    \item \textbf{Random}: Here the allocation of sensing REs is random in the time-frequency domain. As evidenced in Fig. 1(d), this random allocation mitigates range and velocity ambiguities but may lead to an increase in delay-Doppler sidelobe levels.
\end{itemize}

In summary, conventional fixed resource allocation schemes face significant challenges in simultaneously balancing diverse sensing and communication requirements. More importantly, such fixed schemes inadequately exploit the inherent time-frequency selectivity of communication channels, resulting in suboptimal utilization of the limited time-frequency resources, consequently limiting the overall system capacity. These critical shortcomings underscore the need to develop advanced adaptive resource allocation strategies capable of achieving a dynamic equilibrium between diverse sensing requirements and communication demands. To this end, in the following we rigorously derive theoretical metrics for sensing performance, including resolution,  sidelobe level, and sensing SNR. These metrics provide the theoretical foundation for the later development of a sensing-oriented adaptive resource allocation algorithm.

\subsection{Delay and Doppler Resolution}
\subsubsection{Delay resolution} The delay resolution is defined as the $3$dB mainlobe width of the range AF $\chi(\tau, 0)$ \cite{Levanon_book_2004}, \cite{Richards_book_2005}, and quantifies the system's ability to distinguish targets in the delay domain. 
For resource allocations with full subcarrier occupancy, the theoretical delay resolution is given by $1/(N \Delta_{f})$. However, for adaptive resource allocation, the delay resolution will depend on the sensing RE selection vector $\mathbf{u}_0$ and the power allocation vector $\mathbf{p}$. To rigorously characterize this dependence, we generalize the resolution expression as
\begin{equation}
    \Delta \tau = 2(\tau_0 + \delta_{\tau}),
\end{equation}
where $\tau_0 = 1/(2N \Delta_{f})$ represents the baseline for the case of full subcarrier utilization, and $\delta_{\tau} \in \mathbb{R}$ quantifies the resolution deviation induced by adaptive resource distribution. Based on the definition of the $3$dB mainlobe width, the deviation term $\delta_{\tau}$ should satisfy
\begin{equation} \label{eq:defination_delay_3dB}
    \Big| \frac{ \chi(\Delta \tau / 2, 0) }{ \chi(0, 0) } \Big|^2 = \frac{ |\mathbf{u}_0^T (\mathbf{I}_M \otimes \boldsymbol{\Phi}_{\tau_0+\delta_{\tau}})\mathbf{p}|^2 }{ |\mathbf{u}_0^T \mathbf{p}|^2 } = \frac{1}{2}.
\end{equation}
Moreover, $\delta_\tau$ exhibits distinct characteristics depending on the resource allocation: \textit{(i)} For contiguous subcarrier allocation (e.g. FDM), $\delta_\tau >0$ due to reduced effective bandwidth, degrading resolution \cite{Wang_arxiv_2024}. \textit{(ii)} For sparse periphery-focused allocation, $\delta_\tau < 0$ achieves super-resolution at the cost of severe range ambiguities from insufficient sampling \cite{Keskin_TSP_2021}. To derive a closed-form solution for $\delta_{\tau}$, we approximate $\boldsymbol{\Phi}_{\tau_0+\delta_{\tau}}$ using a first-order Taylor expansion as
\begin{equation} \label{eq:delay_steerv_Taylor}
    \boldsymbol{\Phi}_{\tau_0+\delta_{\tau}} \approx \boldsymbol{\Phi}_{\tau_0} - \jmath 2 \pi \Delta_{f} \delta_{\tau} \mathbf{N} \boldsymbol{\Phi}_{\tau_0},
\end{equation}
where we define $\mathbf{N} \triangleq \text{diag}\{0, 1, \dots, N-1\}$. Substituting (\ref{eq:delay_steerv_Taylor}) into (\ref{eq:defination_delay_3dB}) yields
\begin{subequations} \label{eq:delta_tau_expand}
    \begin{align}
    & |\mathbf{u}_0^T (\mathbf{I}_M \otimes \boldsymbol{\Phi}_{\tau_0+\delta_{\tau}})\mathbf{p}|^2   \notag   \\
    \approx &  | \mathbf{u}_0^T (\mathbf{I}_M \otimes \boldsymbol{\Phi}_{\tau_0})  \mathbf{p} - \jmath 2 \pi \Delta_f \delta_{\tau} \mathbf{u}_0^T (\mathbf{I}_M \otimes \mathbf{N} \boldsymbol{\Phi}_{\tau_0}) \mathbf{p} |^2   \\
    =  & | \mathbf{u}_0^T (\mathbf{I}_M \! \otimes \! \boldsymbol{\Phi}_{\tau_0}) \mathbf{p} |^2 \!-\! 4 \pi^2 \Delta_f^2 \delta_{\tau}^2 | \mathbf{u}_0^T (\mathbf{I}_M \! \otimes \! \mathbf{N}\boldsymbol{\Phi}_{\tau_0}) \mathbf{p} |^2  \notag    \\
    & \!-\! 4 \pi \Delta_f \delta_{\tau} \mathfrak{I} \big\{ \mathbf{u}_0^T (\mathbf{I}_M \! \otimes \! \boldsymbol{\Phi}_{\tau_0}) \mathbf{p} \mathbf{p}^T (\mathbf{I}_M \! \otimes \! \boldsymbol{\Phi}_{\tau_0}^H \mathbf{N})\mathbf{u}_0 \big\} \\
    = & | \mathbf{u}_0^T (\mathbf{I}_M \otimes \boldsymbol{\Phi}_{\tau_0}) \mathbf{p} |^2 - 4 \pi^2 \Delta_f^2 \delta_{\tau}^2 | \mathbf{u}_0^T (\mathbf{I}_M \otimes \mathbf{N}\boldsymbol{\Phi}_{\tau_0}) \mathbf{p} |^2 \notag       \\
    & -  4 \pi \Delta_f \delta_{\tau} \mathbf{u}_0^T \mathbf{B} \mathbf{u}_0, \label{eq:Im_delay}
    \end{align}
\end{subequations}
where $\mathbf{B} = \mathfrak{I}\big\{  (\mathbf{I}_M \otimes \boldsymbol{\Phi}_{\tau_0}) \mathbf{p} \mathbf{p}^T (\mathbf{I}_M \otimes \boldsymbol{\Phi}_{\tau_0}^H \mathbf{N})\big\}$, and we employ $\mathfrak{I}\big\{ \mathbf{x}^H \boldsymbol{\Phi} \mathbf{x} \big\} = \mathbf{x}^H \mathfrak{I}\big\{ \boldsymbol{\Phi} \big\} \mathbf{x}$ in (\ref{eq:Im_delay}) since $\mathbf{u}_0$ is a real-valued vector. Given the sufficiently small value of $\delta_{\tau}$, the higher-order terms of $\delta_{\tau}$ in (\ref{eq:delta_tau_expand}) can be disregarded, leading to a simplified solution for $\delta_{\tau}$:
\begin{equation}
    \delta_{\tau} = \frac{ 2| \mathbf{u}_0^T (\mathbf{I}_M \otimes \boldsymbol{\Phi}_{\tau_0}) \mathbf{p} |^2 - |\mathbf{u}_0^T \mathbf{p} |^2 }{ 8 \pi \Delta_f \mathbf{u}_0^T \mathbf{B} \mathbf{u}_0}.
\end{equation}
The corresponding delay resolution is then expressed as
\begin{equation}
    \Delta \tau = \frac{1}{N \Delta_f} + 2\delta_{\tau}. \label{eq:closed_solution_delay_res}
\end{equation}

\subsubsection{Doppler resolution} The Doppler resolution, similarly defined as the $3$dB mainlobe width of the Doppler AF $\chi(0, f_{\text{d}})$, can be expressed as
\begin{equation}
    \Delta f_{\text{d}} = 2(f_0 + \delta_{f_{\text{d}}}),
\end{equation}
where $f_0 = 1/(2 M T_{\text{sym}})$ represents the baseline using the full observation duration. The term $\delta_{f_{\text{d}}} \in \mathbb{R}$ quantifies the Doppler resolution deviation caused by adaptive resource allocation and should satisfy the $3$dB criterion:
\begin{equation} \label{eq:defination_doppler_3dB}
    \Big| \frac{ \chi(0, \Delta f_{\text{d}}/2) }{ \chi(0, 0) } \Big|^2 = \frac{ |\mathbf{u}_0^T (\boldsymbol{\Psi}_{f_0+\delta_{f_{\text{d}}}}^H \otimes \mathbf{I}_N)\mathbf{p}|^2 }{ |\mathbf{u}_0^T \mathbf{p}|^2 } = \frac{1}{2}.
\end{equation}
To derive a closed-form solution for $\delta_{f_{\text{d}}}$, we employ a first-order Taylor expansion to approximate $\boldsymbol{\Psi}_{f_0+\delta_{f_{\text{d}}}}$ as
\begin{equation} \label{eq:doppler_steerv_Taylor}
    \boldsymbol{\Psi}_{f_0+\delta_{f_{\text{d}}}} \approx \boldsymbol{\Psi}_{f_0} - \jmath 2 \pi T_{\text{sym}} \delta_{f_{\text{d}}} \mathbf{M} \boldsymbol{\Psi}_{f_0},
\end{equation}
where $\mathbf{M} \triangleq \text{diag}\{0, 1, \dots, M-1\}$. Substituting (\ref{eq:doppler_steerv_Taylor}) into (\ref{eq:defination_doppler_3dB}), we have
\begin{subequations} \label{eq:delta_doppler_expand}
    \begin{align}
    & |\mathbf{u}_0^T (\boldsymbol{\Psi}_{f_0+\delta_{f_{\text{d}}}}^H \otimes \mathbf{I}_N)\mathbf{p}|^2 \notag  \\
    \approx &|\mathbf{u}_0^T (\boldsymbol{\Psi}_{f_0}^H \otimes \mathbf{I}_N)\mathbf{p} + \jmath 2 \pi T_{\text{sym}} \delta_{f_{\text{d}}} \mathbf{u}_0^T ( \boldsymbol{\Psi}_{f_0}^H \mathbf{M} \!\otimes\! \mathbf{I}_N ) \mathbf{p} |^2  \\
    =  & | \mathbf{u}_0^T (\boldsymbol{\Psi}_{f_0}^H \!\otimes\! \mathbf{I}_N)\mathbf{p} |^2 + 4 \pi^2 T_{\text{sym}}^2 \delta_{f_{\text{d}}}^2 | \mathbf{u}_0^T (\boldsymbol{\Psi}_{f_0}^H \mathbf{M} \!\otimes\! \mathbf{I}_N) \mathbf{p} |^2  \notag \\
    & + 4 \pi T_{\text{sym}} \delta_{f_{\text{d}}} \mathbf{u}_0^T \mathbf{D} \mathbf{u}_0, \label{eq:Im_dopper}
    \end{align}
\end{subequations}
where $\mathbf{D} = \mathfrak{I}\big\{ (\boldsymbol{\Psi}_{f_0}^H \otimes \mathbf{I}_N)\mathbf{p}  \mathbf{p}^T (\mathbf{M} \boldsymbol{\Psi}_{f_0} \otimes \mathbf{I}_N)\big\}$. After neglecting the higher-order terms, closed-form solutions for $\delta_{f_{\text{d}}}$ and Doppler resolution $\Delta f_{\text{d}}$ can be obtained:
\begin{subequations}
    \begin{align}
    \delta_{f_{\text{d}}}   & = \frac{ |\mathbf{u}_0^T \mathbf{p} |^2 - 2| \mathbf{u}_0^T (\boldsymbol{\Psi}_{f_0}^H \otimes \mathbf{I}_N)\mathbf{p} |^2 }{ 8 \pi T_{\text{sym}} \mathbf{u}_0^T \mathbf{D} \mathbf{u}_0 }, \\
    \Delta f_{\text{d}} & = \frac{1}{MT_{\text{sym}}} + 2\delta_{f_{\text{d}}}. \label{eq:closed_solution_doppler_res}
    \end{align}
\end{subequations}

\subsection{Delay-Doppler Sidelobes}
As previously mentioned, the random allocation of sensing REs may lead to high delay-Doppler sidelobe levels. These elevated sidelobes may mask the presence of weak targets, thereby deteriorating the parameter estimation performance. 
To evaluate the delay-Doppler sidelobe level, we sample the AF as
\begin{subequations}
    \begin{align}
        \vartheta(l, \nu)
         & = \chi\left(\frac{l}{N \Delta_f}, \frac{\nu}{M T_{\text{sym}}}\right)  \\
         & = \frac{1}{MN} \sum_{m=0}^{M-1} \sum_{n=0}^{N-1} u_{n, m}^0 p_{n,m} e^{-\jmath 2 \pi l \frac{n}{N}} e^{\jmath 2 \pi \nu \frac{m}{M}} \\
         & = \frac{1}{MN} \mathbf{u}_0 (\boldsymbol{\Psi}_{\nu}^H \otimes \boldsymbol{\Phi}_l) \mathbf{p},
    \end{align}
\end{subequations}
where $l$ and $\nu$ denote the indices of the delay and Doppler bins, respectively, and $\boldsymbol{\Phi}_l \in \mathbb{C}^{N \times N}$ and $\boldsymbol{\Psi}_{\nu} \in \mathbb{C}^{M \times M}$ are diagonal matrices with $\boldsymbol{\Phi}_l(n+1, n+1) = e^{-\jmath 2 \pi l \frac{n}{N}}$ and $\boldsymbol{\Psi}_{\nu}(m+1, m+1) = e^{-\jmath 2 \pi \nu \frac{m}{M}}$. In this paper, we employ the following definition of PSL to quantify the delay-Doppler sidelobe level \cite{Song_TSP_2016}:
\begin{equation}
    \text{PSL} = \underset{ (l, \nu)\in \Omega_{\text{s}}}{\max}  \frac{|\vartheta(l, \nu) |}{MN} = \underset{ (l, \nu) \in \Omega_{\text{s}}}{\max} \frac{|\mathbf{u}_0^T  (\boldsymbol{\Psi}_{\nu}^H \otimes \boldsymbol{\Phi}_l) \mathbf{p} |}{MN},
\end{equation}
where $\Omega_{\text{s}} = \big\{ (l, \nu) \big| ~ |l| \leq l_{\text{max}}, |\nu| \leq \nu_{\text{max}}, (l, \nu) \neq (0, 0) \big\}$ denotes the sidelobe region of interest, and $l_{\text{max}}$ and $\nu_{\text{max}}$ are the maximum values of the delay and Doppler indices, respectively.

\subsection{Sensing SNR}
Since the received echo signal will be distorted by noise, maintaining a sufficient SNR is crucial for achieving satisfactory sensing performance. Given that the target ranges are unknown, we assume the maximum sensing range of interest to be $R_0 = l_{\text{max}} \frac{c}{2B}$. Given the echo signal of the sensing REs in (\ref{eq:echo_sig_sensingRE}), the sensing SNR of a target with maximum range $R_0$ and RCS $\sigma_{\text{rcs}, 0}$ can be written as
\begin{subequations}
    \begin{align}
    \text{SNR}_{\text{r}}& = \frac{ \mathbb{E}\big\{\sum_{m=0}^{M-1} \sum_{n=0}^{N-1} |\alpha_0 u_{n,m}^0 \sqrt{p_{n,m}} s_{n,m}^{\text{r}}|^2 \big\} }{ \mathbb{E}\big\{\sum_{m=0}^{M-1} \sum_{n=0}^{N-1} |u_{n,m}^0 z_{n,m}|^2 \big\} } \\
    & = \frac{ \lambda^2 \mathbb{E}\{\sigma_{\text{rcs,0}}\} }{(4\pi)^3 R_0^4} \frac{\mathbf{u}_0^T \mathbf{p}}{ \mathbf{u}_0^T \mathbf{1}_{MN}  \sigma^2}  \\
    & = \frac{\alpha_{\text{avg}} \mathbf{u}_0^T \mathbf{p}}{\mathbf{u}_0^T \mathbf{1}_{MN}\sigma^2},
    \end{align}
\end{subequations}
where $\alpha_{\text{avg}} \triangleq \frac{\lambda^2}{\varsigma (4\pi)^3  R_0^4} $.

The closed-form sensing performance metrics derived above provide a theoretical foundation for optimizing resource allocation in OFDM-ISAC systems. A key challenge lies in balancing heterogeneous sensing requirements, such as resolution enhancement and sidelobe suppression, while maintaining communication efficiency. In complex multi-target environments, the relative importance of these metrics depends on the spatial distribution of the targets and their RCS characteristics: \textit{(i)} When targets are closely spaced in the delay-Doppler plane, resolution becomes the critical factor. A narrower $3$dB mainlobe width facilitates the discrimination of adjacent targets through improved resolution. \textit{(ii)} For spatially separated targets with significant RCS disparities (e.g., unmanned aerial vehicles near buildings), strong targets generate elevated sidelobes that may mask weaker ones. In this case, sidelobe suppression becomes a priority over resolution enhancement. These contrasting requirements motivate the development of resolution- and sidelobe-oriented adaptive resource allocation strategies, as presented in Sections \ref{sec:min_Res} and \ref{sec:min_PSL}, respectively.

\section{Resolution-Oriented Resource Allocation} \label{sec:min_Res}
In this section, we aim to jointly optimize the resource selection vectors for communication and sensing $\mathbf{u}_{k} = \text{vec}(\mathbf{U}_k)$, $\forall k \in \mathcal{K}=\{0,1,2,\ldots,K\}$, and the power allocation vector $\mathbf{p}$, to maximize the weighted delay-Doppler resolution, while satisfying delay-Doppler PSL, sensing SNR, communication sum-rate, and transmit power requirements, together with the constraints on the resource selection vectors. This adaptive resource allocation design problem can be formulated as
\begin{subequations} \label{eq:problem1}
    \begin{align}
    & \underset{ \{\mathbf{u}_k\}_{k=0}^{K}, \mathbf{p} }{\min} ~ \epsilon_{\tau} \frac{\Delta \tau}{\tau_0} + (1 - \epsilon_{\tau}) \frac{\Delta f_{\text{d}}}{f_0}  \label{eq:prob1_obj}  \\
    & ~~~~ \text{s.t.} \hspace{15pt} \underset{ (l, \nu) \in \Omega_{\text{s}}}{\max} |\mathbf{u}_0^T  (\boldsymbol{\Psi}_{\nu}^H \otimes \boldsymbol{\Phi}_l) \mathbf{p} | \leq \beta_0, \label{eq:prob1_psl_con} \\
    & \hspace{41pt}  \text{SNR}_{\text{r}} \geq \Gamma_0, \label{eq:prob1_snr_con}  \\
    & \hspace{41pt}   R_{\text{c}} \geq \eta_0, \label{eq:prob1_sumrate_con} \\
    & \hspace{41pt} \mathbf{p}^T \mathbf{1}_{MN}   \leq P_{\text{t}}, ~ \mathbf{p} \succeq  \mathbf{0}, \label{eq:prob1_power_con}   \\
    & \hspace{41pt} \sum_{k=0}^{K} \mathbf{u}_k \preceq  \mathbf{1}_{MN},  \label{eq:prob1_orthogRE_con}  \\
    & \hspace{41pt}  \mathbf{u}_k \in \{0, 1 \}^{MN}, ~~ \forall k  \in \mathcal{K}, \label{eq:prob1_bool_con}
    \end{align}
\end{subequations}
where the values of $\tau_0$ and $f_0$ are used to normalize the delay and Doppler resolution which are measured with different units, the parameters $\epsilon_{\tau}$ and $1-\epsilon_{\tau}$ are weighting factors for normalized delay and Doppler resolution, $\beta_0$ is the PSL threshold, $\Gamma_0$ is the sensing SNR threshold, $\eta_0$ is the communication sum-rate threshold, and $P_{\text{t}}$ is the transmit power budget. Clearly, problem (\ref{eq:problem1}) is difficult to solve due to the fractional objective function (\ref{eq:prob1_obj}), the Boolean constraint (\ref{eq:prob1_bool_con}), and various coupled variables. To address these challenges, we will use Dinkelbach's transform and MM to convert (\ref{eq:problem1}) into two tractable sub-problems and iteratively solve them.

\subsection{Dinkelbach's Transform} \label{sec:min_Res_Dinkelbach}
Ignoring the constant terms, the fractional objective function (\ref{eq:prob1_obj}) can be simplified as
\begin{equation} \label{eq:obj_simplify}
    \begin{aligned}
    & \epsilon_\tau \frac{ 2| \mathbf{u}_0^T (\mathbf{I}_M \otimes \boldsymbol{\Phi}_{\tau_0}) \mathbf{p} |^2 - |\mathbf{u}_0^T \mathbf{p} |^2 }{ c_\tau \mathbf{u}_0^T \mathbf{B} \mathbf{u}_0 } + (1-\epsilon_\tau) \\
    & \hspace{2cm} \times \frac{ |\mathbf{u}_0^T \mathbf{p} |^2 - 2| \mathbf{u}_0^T (\boldsymbol{\Psi}_{f_0}^H \otimes \mathbf{I}_N)\mathbf{p} |^2 }{ c_v \mathbf{u}_0^T \mathbf{D} \mathbf{u}_0 },
    \end{aligned}
\end{equation}
where $c_{\tau} = 8 \pi \Delta_f \tau_0$, $c_v = 8 \pi f_0 T_{\text{sym}}$. We should emphasize that the signs of $\mathbf{u}_0^T \mathbf{B} \mathbf{u}_0$ and $\mathbf{u}_0^T \mathbf{D} \mathbf{u}_0$ are indeterminate. To convert the non-convex fractional function (\ref{eq:obj_simplify}) into a polynomial function, we provide the following proposition.

\begin{prop} \label{prop:res_polyno}
    Introducing the two auxiliary variables $t_{\tau} \in \mathbb{R}$ and $t_v \in \mathbb{R}$, the fractional function (\ref{eq:obj_simplify}) becomes
    \begin{equation} \label{eq:obj_poly}
        \epsilon_\tau g_\tau (\mathbf{u}_0, \mathbf{p}) + (1-\epsilon_\tau) g_v (\mathbf{u}_0, \mathbf{p}),
    \end{equation}
    where for clarity we define
    \begin{subequations}
        \begin{align}
             & \hspace{-3mm} g_\tau (\mathbf{u}_0, \mathbf{p}) \!\triangleq\! 2| \mathbf{u}_0^T (\mathbf{I}_M \!\otimes\! \boldsymbol{\Phi}_{\tau_0}) \mathbf{p} |^2 \!-\! |\mathbf{u}_0^T \mathbf{p} |^2 \!+\! c_{\tau} t_\tau \mathbf{u}_0^T \mathbf{B}\mathbf{u}_0, \\
             & \hspace{-3mm} g_v (\mathbf{u}_0, \mathbf{p}) \!\triangleq\! 2| \mathbf{u}_0^T (\boldsymbol{\Psi}_{f_0}^H \!\otimes\! \mathbf{I}_N)\mathbf{p} |^2 \!-\! |\mathbf{u}_0^T \mathbf{p} |^2 \!+\! c_v t_v \mathbf{u}_0^T \mathbf{D} \mathbf{u}_0.
        \end{align}
    \end{subequations}
\end{prop}
\noindent \textit{Proof:} See Appendix \ref{appendix:res_polyno}.
\hfill $\blacksquare$

\vspace{\baselineskip}
 According to Dinkelbach's transform \cite{Dinkelbach_ManageSci_1967}, the auxiliary variables $t_\tau$ and $t_v$ are updated by
\begin{subequations} \label{eq:update_aux}
    \begin{align}
        t_{\tau}^{\star} & = \frac{ 2| \mathbf{u}_0^T (\mathbf{I}_M \otimes \boldsymbol{\Phi}_{\tau_0}) \mathbf{p} |^2 - |\mathbf{u}_0^T \mathbf{p} |^2 }{ c_\tau \mathbf{u}_0^T \mathbf{B} \mathbf{u}_0 }, \\
        t_v^{\star}      & = \frac{ |\mathbf{u}_0^T \mathbf{p} |^2 - 2| \mathbf{u}_0^T (\boldsymbol{\Psi}_{f_0}^H \otimes \mathbf{I}_N)\mathbf{p} |^2 }{ c_v \mathbf{u}_0^T \mathbf{D} \mathbf{u}_0 }.
    \end{align}
\end{subequations}
To further facilitate the development of the algorithm, we reformulate $\mathbf{u}_0^T \mathbf{B} \mathbf{u}_0$ and $\mathbf{u}_0^T \mathbf{D} \mathbf{u}_0$ as
\begin{subequations}
    \begin{align}
        \mathbf{u}_0^T \mathbf{B} \mathbf{u}_0 & = \mathbf{u}_0^T \widetilde{\mathbf{P}} \mathbf{B}_0 \widetilde{\mathbf{P}}\mathbf{u}_0, \\
        \mathbf{u}_0^T \mathbf{D} \mathbf{u}_0 & = \mathbf{u}_0^T \widetilde{\mathbf{P}} \mathbf{D}_0 \widetilde{\mathbf{P}}\mathbf{u}_0,
    \end{align}
\end{subequations}
where 
\begin{equation}
    \begin{aligned}
        \mathbf{n}   & \triangleq [0, 1, \dots, N-1]^T, ~~~ \mathbf{m} \triangleq [0, 1, \dots, M-1]^T,\\
        \mathbf{Q}_1 & \triangleq \mathbf{1}_{M \times M} \otimes \boldsymbol{\phi}(\tau_0) (\mathbf{n}^T \odot \boldsymbol{\phi}^H(\tau_0)),  \\
        \mathbf{Q}_2 & \triangleq \boldsymbol{\psi}^{\ast}(f_0) (\mathbf{m}^T \odot \boldsymbol{\psi}^T(f_0)) \otimes \mathbf{1}_{N \times N},  \\
        \mathbf{B}_0 & \triangleq \frac{\mathbf{Q}_1 - \mathbf{Q}_1^H}{2\jmath}, \hspace{2mm} \mathbf{D}_0 \triangleq \frac{\mathbf{Q}_2 - \mathbf{Q}_2^H}{2\jmath}, \hspace{2mm} \widetilde{\mathbf{P}} \triangleq \text{diag}\{\mathbf{p}\}.
    \end{aligned}
\end{equation}
Clearly, $\mathbf{B}_0$ and $\mathbf{D}_0$ are low-rank Hermitian matrices with zero trace.	Therefore, they can be re-constructed via the eigenvalue decomposition (EVD) as $\mathbf{B}_0 = \mathbf{B}_1 - \mathbf{B}_{-1}$, and $\mathbf{D}_0 = \mathbf{D}_1 - \mathbf{D}_{-1}$, where $\mathbf{B}_1$, $\mathbf{B}_{-1}$, $\mathbf{D}_1$, and $\mathbf{D}_{-1}$ are positive semi-definite matrices. Based on the above derivations, we can transform the non-convex functions in (\ref{eq:obj_poly}) into difference-of-convex (DC) expressions as 
\begin{subequations}\begin{align}
& g_\tau ( \mathbf{u}_0,  \mathbf{p}) = 2| \mathbf{u}_0^T (\mathbf{I}_M \otimes \boldsymbol{\Phi}_{\tau_0}) \mathbf{p} |^2 - |\mathbf{u}_0^T \mathbf{p} |^2 \notag \\
& \hspace{2.3cm} - c_{\tau} t_\tau \mathbf{u}_0^T \widetilde{\mathbf{P}} (\mathbf{B}_1 - \mathbf{B}_{-1}) \widetilde{\mathbf{P}}\mathbf{u}_0, \label{eq:delay_res_DCcon}       \\
& g_v ( \mathbf{u}_0,  \mathbf{p}) = 2| \mathbf{u}_0^T (\boldsymbol{\Psi}_{f_0}^H \otimes \mathbf{I}_N)\mathbf{p} |^2 - |\mathbf{u}_0^T \mathbf{p} |^2 \notag      \\
& \hspace{2.3cm}+ c_v t_v \mathbf{u}_0^T \widetilde{\mathbf{P}} (\mathbf{D}_1 - \mathbf{D}_{-1}) \widetilde{\mathbf{P}}\mathbf{u}_0. \label{eq:dopp_res_DCcon}\end{align}
\end{subequations}

\subsection{Boolean Constraint Relaxation}
To handle the Boolean constraint (\ref{eq:prob1_bool_con}), we incorporate a quadratic penalty term $h(\mathbf{u}_k) = \mathbf{u}_k^T (\mathbf{1}_{MN} -\mathbf{u}_k)$ into the objective function and relax the Boolean constraint to be the box constraint $\mathbf{0}_{MN} \preceq \mathbf{u}_k \preceq \mathbf{1}_{MN}$ \cite{Wang_TSP_2014}, \cite{Liu_TWC_2024}. Hence, the original problem (\ref{eq:problem1}) can be reformulated as
\begin{subequations} \label{eq:prob1_DCform}
    \begin{align}
         & \hspace{-3mm} \underset{\{\mathbf{u}_k\}_{k=0}^K}{\min} \epsilon_\tau g_\tau (\mathbf{u}_0,\! \mathbf{p}) \!+\! (1 \!-\! \epsilon_\tau) g_v (\mathbf{u}_0,\! \mathbf{p}) \!+\! \rho \! \sum_{k=0}^{K} \! h(\mathbf{u}_k) \! \label{eq:prob1_DCform_obj} \\
         & \hspace{1pt} \text{s.t.} \hspace{9pt} \text{(\ref{eq:prob1_psl_con})}-\text{(\ref{eq:prob1_orthogRE_con})},  \\
         & \hspace{22pt} \mathbf{0}_{MN} \preceq \mathbf{u}_k \preceq \mathbf{1}_{MN}, ~\forall k \in \mathcal{K}. \label{eq:prob1_DCfrom_box_con}
    \end{align}
\end{subequations}
Problem (\ref{eq:prob1_DCform}) consists of a DC objective function (\ref{eq:prob1_DCform_obj}), a second-order cone constraint (\ref{eq:prob1_psl_con}), and linear constraints (\ref{eq:prob1_snr_con})-(\ref{eq:prob1_orthogRE_con}), (\ref{eq:prob1_DCfrom_box_con}). Hence, the non-convexity of (\ref{eq:prob1_DCform}) stems from the DC functions and coupled variables. In the following, we decompose~(\ref{eq:prob1_DCform}) into two sub-problems and alternately optimize them using MM \cite{SunY_TSP_2017}.

\subsection{MM-based Update}
\subsubsection{Update $\mathbf{u}_k$} Given the power allocation vector $\mathbf{p}^{(i)}$ and auxiliary variables $t_{\tau}^{(i)}$ and $t_v^{(i)}$ in the $i$-th iteration, the sub-problem for optimizing the resource selection vector $\mathbf{u}_k$ is
\begin{subequations} \label{eq:prob1_update_u}
    \begin{align}
         & \underset{\{\mathbf{u}_k\}_{k=0}^K}{\min} ~ \epsilon_\tau g_\tau (\mathbf{u}_0) \!+\! (1 \!-\! \epsilon_\tau) g_v (\mathbf{u}_0) \!+\! \rho \! \sum_{k=0}^{K} \! h(\mathbf{u}_k)  \!\!  \label{eq:prob1_update_u_obj} \\
         & \hspace{9pt} \text{s.t.} \hspace{14pt} \text{(\ref{eq:prob1_psl_con})}-\text{(\ref{eq:prob1_sumrate_con})}, ~ \text{(\ref{eq:prob1_orthogRE_con})}, ~\text{(\ref{eq:prob1_DCfrom_box_con})}.
    \end{align}
\end{subequations}
Sub-problem (\ref{eq:prob1_update_u}) is a DC programming problem and can be solved by applying the MM method. We observe that the objective function in (\ref{eq:prob1_update_u_obj}) is composed of multiple DC functions. Generally, the surrogate for the DC function can be constructed by linearizing its concave components. However, since the signs of $t_\tau$ and $t_v$ are indeterminate, we need to consider different cases for the concave terms in $g_\tau(\mathbf{u}_0)$ and $g_v (\mathbf{u}_0)$ depending on the signs of $t_\tau$ and $t_v$. To express this more succinctly, we define $r = \text{sign}(t_\tau)$ and $j = -\text{sign}(t_v)$. With these definitions, the concave parts of $g_\tau (\mathbf{u}_0)$ and $g_v (\mathbf{u}_0)$ are given by $-\mathbf{u}_0^T \widetilde{\mathbf{P}} \mathbf{B}_r \widetilde{\mathbf{P}}\mathbf{u}_0$ and $-\mathbf{u}_0^T \widetilde{\mathbf{P}} \mathbf{D}_j \widetilde{\mathbf{P}}\mathbf{u}_0$, respectively. Thus, we can linearize the concave terms of the DC functions using a first-order Taylor expansion:
\begin{subequations} \label{eq:surrogate_func}
    \begin{align}
    - \|\mathbf{u}_k\|^2 & \leq -2 \mathbf{u}_k^T \mathbf{u}_k^{(i)} + \| \mathbf{u}_k^{(i)}\|^2, \\
    -|\mathbf{u}_0^T \mathbf{p}|^2 & \leq -2 \mathbf{u}_0^T \mathbf{p} \mathbf{p}^T \mathbf{u}_0^{(i)} + |\mathbf{p}^T \mathbf{u}_0^{(i)}|^2,  \\
    \!\!\!\!\! -\mathbf{u}_0^T \widetilde{\mathbf{P}} \mathbf{B}_r \widetilde{\mathbf{P}}\mathbf{u}_0 & \leq -2 \mathbf{u}_0^T \widetilde{\mathbf{P}} \mathbf{B}_r \widetilde{\mathbf{P}}\mathbf{u}_0^{(i)} \!\!+\! (\mathbf{u}_0^{(i)})^T \widetilde{\mathbf{P}} \mathbf{B}_r \widetilde{\mathbf{P}}\mathbf{u}_0^{(i)}, \\
    \!\!\!\!\! -\mathbf{u}_0^T \widetilde{\mathbf{P}} \mathbf{D}_j \widetilde{\mathbf{P}}\mathbf{u}_0 & \leq -2 \mathbf{u}_0^T \widetilde{\mathbf{P}} \mathbf{D}_j \widetilde{\mathbf{P}}\mathbf{u}_0^{(i)} \!\!+\! (\mathbf{u}_0^{(i)})^T \widetilde{\mathbf{P}} \mathbf{D}_j \widetilde{\mathbf{P}}\mathbf{u}_0^{(i)},
    \end{align}
\end{subequations}
where $\mathbf{u}_k^{(i)}$ denotes the value of $\mathbf{u}_k$ in the $i$-th iteration. Then, the surrogate function for the DC objective is written as
\begin{equation}
    \epsilon_\tau g_\tau^{(i)} (\mathbf{u}_0) + (1-\epsilon_\tau) g_v^{(i)} (\mathbf{u}_0) + \rho \sum_{k=0}^{K} h^{(i)}(\mathbf{u}_k),
\end{equation}
where for brevity we define
\begin{subequations}
    \begin{align}
    g_\tau^{(i)} (\mathbf{u}_0) & \triangleq 2| \mathbf{u}_0^T (\mathbf{I}_M \otimes\boldsymbol{\Phi}_{\tau_0}) \mathbf{p} |^2 + c_{\tau} |t_\tau| \mathbf{u}_0^T \widetilde{\mathbf{P}} \mathbf{B}_{-r} \widetilde{\mathbf{P}}\mathbf{u}_0 \notag \\
    & \hspace{3.4cm} - \mathbf{u}_0^T \mathbf{b}_r^{(i)}  + c_1^{(i)},  \\
    g_v^{(i)} (\mathbf{u}_0) & \triangleq 2| \mathbf{u}_0^T (\boldsymbol{\Psi}^H_{f_0} \otimes \mathbf{I}_N) \mathbf{p} |^2 + c_v |t_v| \mathbf{u}_0^T \widetilde{\mathbf{P}} \mathbf{D}_{-j} \widetilde{\mathbf{P}}\mathbf{u}_0 \notag \\
    & \hspace{3.4cm} - \mathbf{u}_0^T \mathbf{d}_j^{(i)} + c_2^{(i)}, \\
    h^{(i)}(\mathbf{u}_k)   & \triangleq \mathbf{u}_k^T (\mathbf{1}_{MN} - 2 \mathbf{u}_k^{(i)}) +  \| \mathbf{u}_k^{(i)} \|^2 , \label{eq:h_cvx}  \\
    \mathbf{b}_r^{(i)}  & \triangleq 2\mathbf{p} \mathbf{p}^T \mathbf{u}_0^{(i)} + 2 c_{\tau} |t_\tau| \widetilde{\mathbf{P}} \mathbf{B}_r \widetilde{\mathbf{P}}\mathbf{u}_0^{(i)},  \\
    \mathbf{d}_j^{(i)}  & \triangleq 2\mathbf{p} \mathbf{p}^T \mathbf{u}_0^{(i)} + 2 c_v |t_v| \widetilde{\mathbf{P}} \mathbf{D}_j \widetilde{\mathbf{P}}\mathbf{u}_0^{(i)}, \\
    c_1^{(i)}    & \triangleq | \mathbf{p}^T \mathbf{u}_0^{(i)} |^2 + c_{\tau} |t_\tau| (\mathbf{u}_0^{(i)})^T \widetilde{\mathbf{P}} \mathbf{B}_r \widetilde{\mathbf{P}}\mathbf{u}_0^{(i)},  \\
    c_2^{(i)}  & \triangleq | \mathbf{p}^T \mathbf{u}_0^{(i)} |^2 + c_v |t_v| (\mathbf{u}_0^{(i)})^T \widetilde{\mathbf{P}} \mathbf{D}_j \widetilde{\mathbf{P}}\mathbf{u}_0^{(i)}.
    \end{align}
\end{subequations}

Based on the above, the update for $\mathbf{u}_k$ is formulated as
\begin{subequations} \label{eq:prob1_update_u_qcqp}
    \begin{align}
    & \underset{\{\mathbf{u}_k\}_{k=0}^K}{\min} ~ \epsilon_\tau g_\tau^{(i)} (\mathbf{u}_0) \!+\! (1\!-\!\epsilon_\tau) g_v^{(i)} (\mathbf{u}_0) \!+\! \rho \sum_{k=0}^{K} h^{(i)}(\mathbf{u}_k) \! \\
    & \hspace{9pt} \text{s.t.} \hspace{14pt} \text{(\ref{eq:prob1_psl_con})}-\text{(\ref{eq:prob1_sumrate_con})}, ~ \text{(\ref{eq:prob1_orthogRE_con})}, ~\text{(\ref{eq:prob1_DCfrom_box_con})},
    \end{align}
\end{subequations}
which is a convex quadratically constrained quadratic program (QCQP) that can be easily solved.

\subsubsection{Update $\mathbf{p}$} After obtaining $\mathbf{u}_k^{(i)}$, $t_{\tau}^{(i)}$, and $t_v^{(i)}$, the optimization for updating the power allocation vector $\mathbf{p}$ can be formulated as
\begin{subequations} \label{eq:prob1_update_p}
    \begin{align}
    & \underset{\mathbf{p}}{\min} ~~ \epsilon_\tau g_\tau (\mathbf{p}) + (1-\epsilon_\tau) g_v (\mathbf{p}) \label{eq:prob1_update_p_obj} \\
    & \hspace{5pt} \text{s.t.} \hspace{10pt} \text{(\ref{eq:prob1_psl_con})}-\text{(\ref{eq:prob1_power_con})},
    \end{align}
\end{subequations}
which is also a DC programming problem. Using derivations similar to (\ref{eq:surrogate_func}), the surrogate function for the DC objective function in(\ref{eq:prob1_update_p_obj}) at the current point $\mathbf{p}^{(i)}$ is given by
\begin{equation}
    \epsilon_\tau {g}_\tau^{(i)} (\mathbf{p}) + (1-\epsilon_\tau) {g}_v^{(i)} (\mathbf{p}),
\end{equation}
where
\begin{subequations} \label{eq:cvx_fun_p}
    \begin{align}
    g_\tau^{(i)} (\mathbf{p}) & = 2| \mathbf{u}_0^T (\mathbf{I}_M \otimes \boldsymbol{\Phi}_{\tau_0}) \mathbf{p} |^2 + c_{\tau} |t_\tau| \mathbf{p}^T \widetilde{\mathbf{U}}_0 \mathbf{B}_{-r} \widetilde{\mathbf{U}}_0 \mathbf{p}  \notag \\
        & \hspace{3.4cm} - \mathbf{p}^T \mathbf{e}_r^{(i)} + c_3^{(i)},\\
    g_v^{(i)} (\mathbf{p})  & = 2| \mathbf{u}_0^T (\boldsymbol{\Psi}^H_{f_0} \otimes \mathbf{I}_N) \mathbf{p} |^2 + c_v |t_v| \mathbf{p}^T \widetilde{\mathbf{U}}_0 \mathbf{D}_{-j} \widetilde{\mathbf{U}}_0 \mathbf{p} \notag \\
    & \hspace{3.4cm} - \mathbf{p}^T \mathbf{q}_j^{(i)} + c_4^{(i)},  \\
    \mathbf{e}_r^{(i)} & = 2 \mathbf{u}_0\mathbf{u}_0^T \mathbf{p}^{(i)} + 2 c_{\tau} |t_\tau| \widetilde{\mathbf{U}}_0 \mathbf{B}_r \widetilde{\mathbf{U}}_0 \mathbf{p}^{(i)}, \\
    \mathbf{q}_j^{(i)} & = 2 \mathbf{u}_0\mathbf{u}_0^T \mathbf{p}^{(i)} + 2 c_v |t_v| \widetilde{\mathbf{U}}_0 \mathbf{D}_j \widetilde{\mathbf{U}}_0 \mathbf{p}^{(i)},
    \end{align}
\end{subequations}
$\widetilde{\mathbf{U}}_0 = \text{diag}\{ \mathbf{u}_0 \}$, and constant terms $c_3^{(i)}$ and $c_4^{(i)}$ are irrelevant to the variable $\mathbf{p}$. Then, the sub-problem for optimizing $\mathbf{p}$ can be expressed as
\begin{subequations} \label{eq:prob1_update_p_qcqp}
    \begin{align}
         & \underset{\mathbf{p}}{\min} ~~  \epsilon_\tau g_\tau^{(i)} (\mathbf{p}) + (1-\epsilon_\tau) g_v^{(i)} (\mathbf{p}) \\
         & \hspace{5pt} \text{s.t.} \hspace{10pt} \text{(\ref{eq:prob1_psl_con})}-\text{(\ref{eq:prob1_power_con})},
    \end{align}
\end{subequations}
which is a convex QCQP that can also be easily solved.

\begin{algorithm}[!t]
    \begin{small}
        \caption{Resolution-Oriented Adaptive Resource Allocation}
        \label{Alg:minRes}
        \begin{algorithmic}[1]
            \REQUIRE {$\mathbf{H}_k$, $\forall k$, $\tau_0$, $f_{\text{d},0}$, $\epsilon_\tau$, $\beta_0$, $\Gamma_0$, $\eta_0$, $\Omega_{\text{s}}$, $\sigma^2$,  $\rho$, $P_{\text{t}}$, $\delta_{\text{th}}$.}
            \ENSURE {$\mathbf{u}_k^\star$, $\mathbf{p}^\star$}
            \STATE {Initialize $\mathbf{u}_k^{(0)} = \mathbf{1}_{MN}/2$, $\forall k$, $\mathbf{p}^{(0)} = P_{\text{t}}/(MN) \mathbf{1}_{MN}$, $i:=0$.}
            \STATE {Calculate the objective value $f_{\text{obj}}^{(0)}$ using (\ref{eq:prob1_DCform_obj}).}
            \REPEAT
            \STATE {Update $\mathbf{u}_{k}^{i+1}$ by solving (\ref{eq:prob1_update_u_qcqp}).}
            \STATE {Update $\mathbf{p}^{i+1}$ by solving (\ref{eq:prob1_update_p_qcqp}).}
            \STATE {Update $t_{\tau}^{i+1}$ and $t_v^{i+1}$ using (\ref{eq:update_aux}).}
            \STATE {Calculate the objective value $f_{\text{obj}}^{(i+1)}$ using (\ref{eq:prob1_DCform_obj}).}
            \STATE {$i:=i+1$.}
            \UNTIL {$|(f_{\text{obj}}^{(i)} - f_{\text{obj}}^{(i+1)}) / f_{\text{obj}}^{(i)} |  \leq \delta_{\text{th}}$}
            \STATE {Return $\mathbf{u}_k^\star = \mathbf{u}_{k}^{(i)}$, $\mathbf{p}^\star = \mathbf{p}^{(i)}$.}
        \end{algorithmic}
    \end{small}
\end{algorithm}

\subsection{Summary and Computational Complexity Analysis}
Based on the above derivations, the proposed resolution-oriented adaptive resource allocation is summarized in Algorithm \ref{Alg:minRes}, where $\delta_{\text{th}}$ represents the convergence threshold. The resolution-oriented optimization in~(\ref{eq:problem1}) is solved by alternately updating $\mathbf{u}_k$, $\mathbf{p}$, $t_\tau$, and $t_v$. The computational complexity of Algorithm \ref{Alg:minRes} is mainly due to the iterative updates of $\mathbf{u}_k$ and $\mathbf{p}$. Assuming the commonly used interior point method is employed to solve these convex QCQP problems, the computational complexity of solving (\ref{eq:prob1_update_u_qcqp}) and $(\ref{eq:prob1_update_p_qcqp})$ is of order $\mathcal{O} \big\{ (MN)^{3.5} (K+1)^{3.5} \big\}$ and $\mathcal{O} \big\{(MN)^{3.5}\}$, respectively. Therefore, the overall computational complexity of Algorithm \ref{Alg:minRes} is of order $\mathcal{O} \big\{ \operatorname{ln}(1/\delta_{\text{th}}) (MN)^{3.5} (K+1)^{3.5} \big\}$.

\section{Sidelobe-Oriented Resource Allocation} \label{sec:min_PSL}
In this section, we consider sidelobe-oriented adaptive resource allocation. In particular, we propose to jointly design the resource selection vectors $\mathbf{u}_{k}$ and the power allocation vector $\mathbf{p}$ to minimize the delay-Doppler PSL while satisfying constraints on the delay and Doppler resolution, sensing SNR, communication sum-rate, transmit power budget, and resource selection vectors. The corresponding optimization problem is formulated as
\begin{subequations} \label{eq:problem2}
    \begin{align}
         & \underset{ \{\mathbf{u}_k\}_{k=0}^K, \mathbf{p} }{\min} ~~ \underset{ (l, \nu) \in \Omega_{\text{s}}}{\max}~ |\mathbf{u}_0^T  (\boldsymbol{\Psi}_{\nu}^H \otimes \boldsymbol{\Phi}_l) \mathbf{p} |  \label{eq:prob2_psl_obj} \\
         & \hspace{14pt} \text{s.t.} \hspace{20pt} \Delta \tau  \leq \tau_{\text{th}} , \label{eq:prob2_delay_res_con} \\
         & \hspace{46pt} \Delta f_{\text{d}}  \leq f_{\text{th}} , \label{eq:prob2_doppler_res_con} \\
         & \hspace{46pt}  \text{(\ref{eq:prob1_snr_con})}-\text{(\ref{eq:prob1_bool_con})},
    \end{align}
\end{subequations}
where $\tau_{\text{th}}$ and $f_{\text{th}}$ are the delay and Doppler resolution thresholds, respectively. Problem (\ref{eq:problem2}) is challenging to solve due to the non-convex fractional constraints in (\ref{eq:prob2_delay_res_con}) and (\ref{eq:prob2_doppler_res_con}), the Boolean constrain (\ref{eq:prob1_bool_con}), and coupled variables. Next, we use transformations similar to those in Sec.~\ref{sec:min_Res} to iteratively solve (\ref{eq:problem2}).

\subsection{Problem Reformulation}
We first use derivations similar to those in Sec.~\ref{sec:min_Res_Dinkelbach} to transform the non-convex fractional constraints (\ref{eq:prob2_delay_res_con}) and (\ref{eq:prob2_doppler_res_con}) into DC constraints as
\begin{subequations}
    \begin{align}
         & \bar{g}_\tau ( \mathbf{u}_0,  \mathbf{p}) = 2| \mathbf{u}_0^T (\mathbf{I}_M \otimes \boldsymbol{\Phi}_{\tau_0}) \mathbf{p} |^2 - |\mathbf{u}_0^T \mathbf{p} |^2 \notag      \\
         & \hspace{2.3cm} - \bar{t}_\tau \mathbf{u}_0^T \widetilde{\mathbf{P}} (\mathbf{B}_1 - \mathbf{B}_{-1}) \widetilde{\mathbf{P}}\mathbf{u}_0 \leq 0,  \\
         & \bar{g}_v( \mathbf{u}_0,  \mathbf{p}) = 2| \mathbf{u}_0^T (\boldsymbol{\Psi}_{f_0}^H \otimes \mathbf{I}_N)\mathbf{p} |^2 - |\mathbf{u}_0^T \mathbf{p} |^2 \notag           \\
         & \hspace{2.3cm}+ \bar{t}_v \mathbf{u}_0^T \widetilde{\mathbf{P}} (\mathbf{D}_1 - \mathbf{D}_{-1}) \widetilde{\mathbf{P}}\mathbf{u}_0 \leq 0, \label{eq:prob2_dopp_res_DCcon}
    \end{align}
\end{subequations}
where $\bar{t}_\tau = 4 \pi (\Delta_f \tau_\text{th} - 1/N)$ and $\bar{t}_v = 4 \pi (f_{\text{th}} T_{\text{sym}} - 1/M)$. Then, we convert the Boolean constraint (\ref{eq:prob1_bool_con}) into a quadratic penalty term $h(\mathbf{u}_k)$ with a box constraint. Problem (\ref{eq:problem2}) thus becomes
\begin{subequations} \label{eq:prob2_DC}
    \begin{align}
         & \underset{ \{\mathbf{u}_k\}_{k=0}^K, \mathbf{p} }{\min} ~~ \underset{ (l, \nu) \in \Omega_{\text{s}}}{\max}~ |\mathbf{u}_0^T  (\boldsymbol{\Psi}_{\nu}^H \otimes \boldsymbol{\Phi}_l) \mathbf{p} | + \rho \sum_{k=0}^{K} h(\mathbf{u}_k) \label{eq:prob2_DC_obj} \\
         & \hspace{15pt} \text{s.t.} \hspace{20pt} \bar{g}_\tau ( \mathbf{u}_0,  \mathbf{p}) \leq 0, \label{eq:prob2_delay_res_DCcon} \\
         & \hspace{47pt}  \bar{g}_v ( \mathbf{u}_0,  \mathbf{p}) \leq 0 , \label{eq:prob2_doppler_res_DCcon} \\
         & \hspace{47pt}  \text{(\ref{eq:prob1_snr_con})}-\text{(\ref{eq:prob1_orthogRE_con})}, ~\text{(\ref{eq:prob1_DCfrom_box_con})}.
    \end{align}
\end{subequations}
As before, this problem can also be decomposed into two sub-problems and alternately solved using the MM method.

\subsection{MM-based Update}
\subsubsection{Update $\mathbf{u}_k$} With fixed $\mathbf{p}$, the optimization for the resource selection vectors $\mathbf{u}_k$ can be formulated as the following DC programming problem:
\begin{subequations} \label{eq:prob2_update_u}
    \begin{align}
         & \underset{\{\mathbf{u}_k\}_{k=0}^K}{\min} ~~ \underset{ (l, \nu) \in \Omega_{\text{s}}}{\max} ~|\mathbf{u}_0^T  (\boldsymbol{\Psi}_{\nu}^H \otimes \boldsymbol{\Phi}_l) \mathbf{p} | + \rho \sum_{k=0}^{K} h(\mathbf{u}_k) \\
         & \hspace{10pt} \text{s.t.} \hspace{18pt}  \bar{g}_\tau (\mathbf{u}_0) \leq 0, \\
         & \hspace{40pt}  \bar{g}_v (\mathbf{u}_0) \leq 0 , \\
         & \hspace{40pt}  \text{(\ref{eq:prob1_snr_con})}, ~\text{(\ref{eq:prob1_sumrate_con})}, ~\text{(\ref{eq:prob1_orthogRE_con})}, ~\text{(\ref{eq:prob1_DCfrom_box_con})}.
    \end{align}
\end{subequations}
We observe that the DC functions $h(\mathbf{u}_k)$, $\bar{g}_\tau (\mathbf{u}_0)$ and $\bar{g}_v (\mathbf{u}_0)$ in problem (\ref{eq:prob2_update_u}) exhibit a structure similar to those in (\ref{eq:prob1_update_u}). Consequently, analogous derivations can be used to transform problem (\ref{eq:prob2_update_u}) as
\begin{subequations} \label{eq:prob2_update_u_qcqp}
    \begin{align}
         & \underset{\{\mathbf{u}_k\}_{k=0}^K}{\min} ~ ~\underset{ (l, \nu) \in \Omega_{\text{s}}}{\max}~ |\mathbf{u}_0^T  (\boldsymbol{\Psi}_{\nu}^H \otimes \boldsymbol{\Phi}_l) \mathbf{p} | + \rho \sum_{k=0}^{K} h^{(i)}(\mathbf{u}_k) \\
         & \hspace{10pt} \text{s.t.} \hspace{18pt}  \bar{g}_\tau^{(i)} (\mathbf{u}_0) \leq 0, \\
         & \hspace{40pt}  \bar{g}_v^{(i)} (\mathbf{u}_0) \leq 0 , \\
         & \hspace{40pt}  \text{(\ref{eq:prob1_snr_con})}, ~\text{(\ref{eq:prob1_sumrate_con})}, ~\text{(\ref{eq:prob1_orthogRE_con})}, ~\text{(\ref{eq:prob1_DCfrom_box_con})},
    \end{align}
\end{subequations}
where $h^{(i)}(\mathbf{u}_k)$ is given in (\ref{eq:h_cvx}). The convex surrogate functions $\bar{g}_\tau^{(i)} (\mathbf{u}_0)$ and $\bar{g}_v^{(i)}(\mathbf{u}_0)$ in the $i$-th iteration can be derived using (\ref{eq:surrogate_func}), but we omit the derivations here. Problem~(\ref{eq:prob2_update_u_qcqp}) is a convex QCQP and thus can be solved efficiently using standard convex tools.

\subsubsection{Update $\mathbf{p}$} Given $\mathbf{u}_k$, the optimization for updating $\mathbf{p}$ is 
\begin{subequations} \label{eq:prob2_update_p}
    \begin{align}
         & \underset{\mathbf{p}}{\min} ~~ \underset{ (l, \nu) \in \Omega_{\text{s}}}{\max} ~|\mathbf{u}_0^T  (\boldsymbol{\Psi}_{\nu}^H \otimes \boldsymbol{\Phi}_l) \mathbf{p} | \\
         & ~\text{s.t.} \hspace{10pt}  \bar{g}_\tau (\mathbf{p}) \leq 0,      \\
         & \hspace{25pt}  \bar{g}_v (\mathbf{p}) \leq 0 ,  \\
         & \hspace{25pt}  \text{(\ref{eq:prob1_snr_con})}-\text{(\ref{eq:prob1_power_con})}.
    \end{align}
\end{subequations}
As before, we employ MM to transform the DC programming problem (\ref{eq:prob2_update_p}) into the following QCQP problem:
\begin{subequations} \label{eq:prob2_update_p_qcqp}
    \begin{align}
         & \underset{\mathbf{p}}{\min} ~~ \underset{ (l, \nu) \in \Omega_{\text{s}}}{\max}~ |\mathbf{u}_0^T  (\boldsymbol{\Psi}_{\nu}^H \otimes \boldsymbol{\Phi}_l) \mathbf{p} | \\
         & ~\text{s.t.} \hspace{10pt}  \bar{g}_\tau^{(i)} (\mathbf{p}) \leq 0,   \\
         & \hspace{25pt}  \bar{g}_v^{(i)} (\mathbf{p}) \leq 0 ,\\
         & \hspace{25pt}  \text{(\ref{eq:prob1_snr_con})}-\text{(\ref{eq:prob1_power_con})},
    \end{align}
\end{subequations}
where $\bar{g}_\tau^{(i)} (\mathbf{p})$ and $\bar{g}_v^{(i)} (\mathbf{p})$ can be derived based on the results in (\ref{eq:cvx_fun_p}). Problem (\ref{eq:prob2_update_u_qcqp}) is convex and thus can be readily solved.

\begin{algorithm}[!t]
    \begin{small}
        \caption{Sidelobe-Oriented Adaptive Resource Allocation}
        \label{Alg:minPSL}
        \begin{algorithmic}[1]
            \REQUIRE {$\mathbf{H}_k$, $\forall k$, $\tau_0$, $f_{\text{d},0}$, $\tau_{\text{th}}$, $f_{\text{th}}$, $\Gamma_0$, $\eta_0$, $\Omega_{\text{s}}$, $\sigma^2$,  $\rho$, $P_{\text{t}}$, $\delta_{\text{th}}$.}
            \ENSURE {$\mathbf{u}_k^\star$, $\mathbf{p}^\star$}
            \STATE {Initialize $\mathbf{u}_k^{(0)} = \mathbf{1}_{MN}/2$, $\forall k$, $\mathbf{p}^{(0)} = P_{\text{t}}/(MN) \mathbf{1}_{MN}$, $i:=0$.}
            \STATE {Calculate the objective value $f_{\text{obj}}^{(0)}$ using (\ref{eq:prob2_DC_obj}).}
            \REPEAT
            \STATE {Update $\mathbf{u}_{k}^{i+1}$ by solving (\ref{eq:prob2_update_u_qcqp}).}
            \STATE {Update $\mathbf{p}^{i+1}$ by solving (\ref{eq:prob2_update_p_qcqp}).}
            \STATE {Calculate the objective value $f_{\text{obj}}^{(i+1)}$ using (\ref{eq:prob2_DC_obj}).}
            \STATE {$i:=i+1$.}
            \UNTIL {$|(f_{\text{obj}}^{(i)} - f_{\text{obj}}^{(i+1)}) / f_{\text{obj}}^{(i)} |  \leq \delta_{\text{th}}$}
            \STATE {Return $\mathbf{u}_k^\star = \mathbf{u}_{k}^{(i)}$, $\mathbf{p}^\star = \mathbf{p}^{(i)}$.}
        \end{algorithmic}
    \end{small}
\end{algorithm}

\subsection{Summary and Computational Complexity Analysis}
The proposed sidelobe-oriented adaptive resource allocation is summarized in Algorithm \ref{Alg:minPSL}.  The update for $\mathbf{u}_k$ and $\mathbf{p}$ dominate the computational burden of Algorithm \ref{Alg:minPSL}, and the complexity of these updates are of order $\mathcal{O} \big\{ (MN)^{3.5} (K+1)^{3.5} \big\}$ and $\mathcal{O} \big\{ (MN)^{3.5} \big\}$, respectively. Thus, the overall complexity of Algorithm \ref{Alg:minPSL} can be approximated as $\mathcal{O} \big\{ \operatorname{ln}(1/\delta_{\text{th}}) (MN)^{3.5} (K+1)^{3.5} \big\}$.

\setlength{\tabcolsep}{0.3 pt}\begin{table}[!t]
    \centering
    \caption{Simulation Parameters}\label{table1}
    {\footnotesize{
            \begin{tabular}{|l|l|l|}
                \hline
                ~\textbf{Parameter }  \hspace{3.2cm} & ~{\textbf{Symbol}}\hspace{4pt} & ~{\textbf{Value}} \hspace{20pt} \\
                \hline
                ~Carrier frequency~                  & ~$f_{\text{c}}$                & ~$28$GHz                        \\
                \hline
                ~Subcarrier spacing~                 & ~$\Delta_f$                    & ~$120$kHz                       \\
                \hline
                ~Number of subcarriers~              & ~$N$                           & ~$256$                          \\
                \hline
                ~Number of symbols~                  & ~$M$                           & ~$128$                          \\
                \hline
                ~Weighting factor~                   & ~$\epsilon_\tau$               & ~$0.5$                          \\
                \hline
                ~Noise PSD~                          & ~$N_0$                         & ~$-150$dBm/Hz                      \\
                \hline
                ~PSL threshold~                      & ~$\beta_0$                     & ~$-40$dB                        \\
                \hline
                ~Sensing SNR threshold~              & ~$\Gamma_0$                    & ~$-10$dB                        \\
                \hline
                ~Communication sum-rate threshold    & ~$\eta_0$                      & ~$3$bps/Hz                      \\
                \hline
                ~The convergence threshold           & ~$\delta_{\text{th}}$          & ~$10^{-4}$                      \\
                \hline
            \end{tabular}
        }}
\end{table}

\begin{figure}[!t]
    \centering
    \includegraphics[width = 3.0 in]{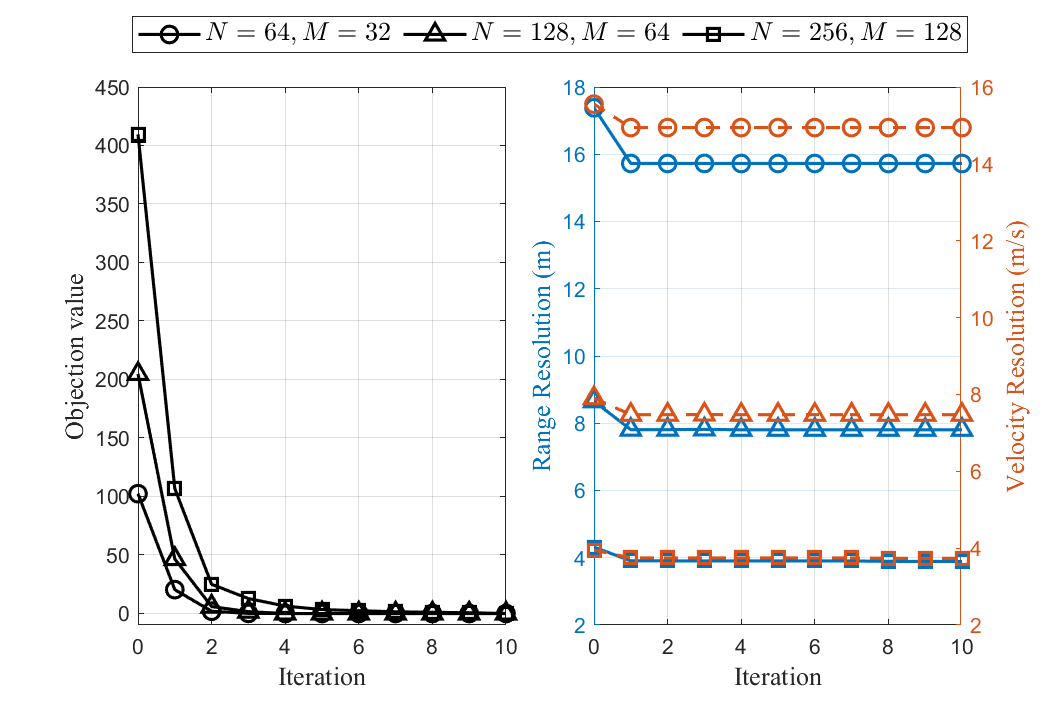}
    \caption{Convergence behavior of Algorithm \ref{Alg:minRes}.}
    \label{fig:converge_minRes}
    \vspace{-2mm}
\end{figure}

\section{Simulation Results}
In this section, we present comprehensive simulation results to evaluate the performance of the proposed resolution-oriented (Algorithm \ref{Alg:minRes}) and sidelobe-oriented (Algorithm \ref{Alg:minPSL}) adaptive resource allocation designs. The default simulation parameters are provided in Table \ref{table1}.

\begin{figure*}[!t]
    \centering
    \subfigbottomskip=-4pt
    \subfigcapskip=-2pt
    \subfigure[$l_{\text{max}} = 0$, $\nu_{\text{max}} = 0$.]{
        \includegraphics[width=0.2\linewidth]{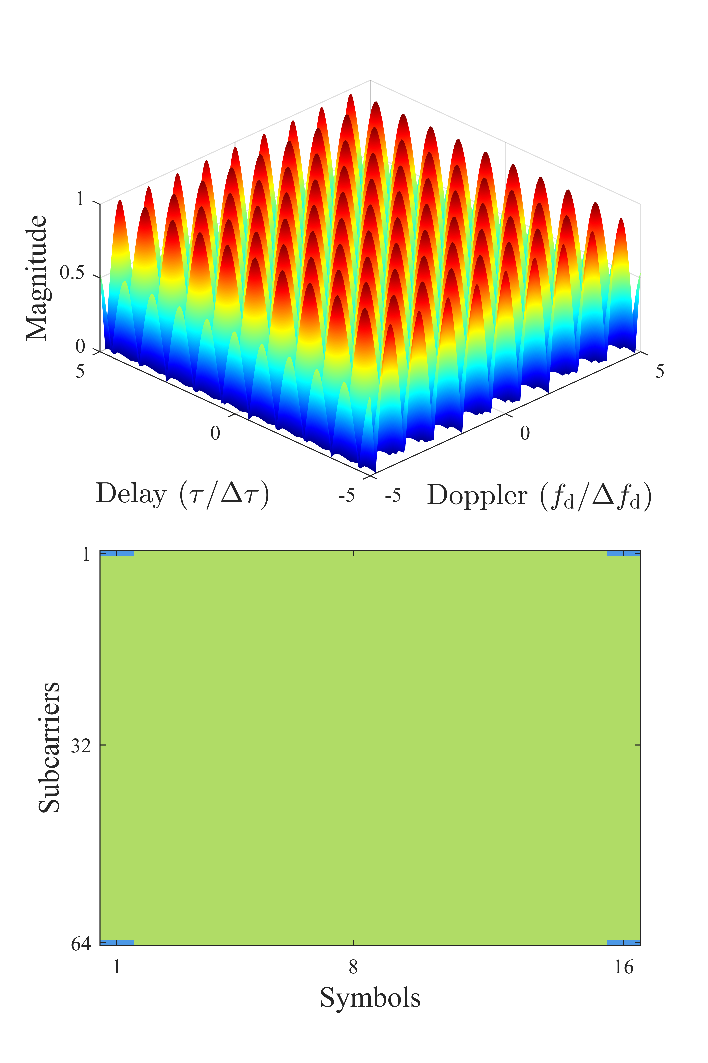} \label{fig:AF_RE_woPSL}
    }
    \hspace{-4mm}
    \subfigure[$l_{\text{max}} = 16$, $\nu_{\text{max}} = 4$.]{
        \includegraphics[width=0.2\linewidth]{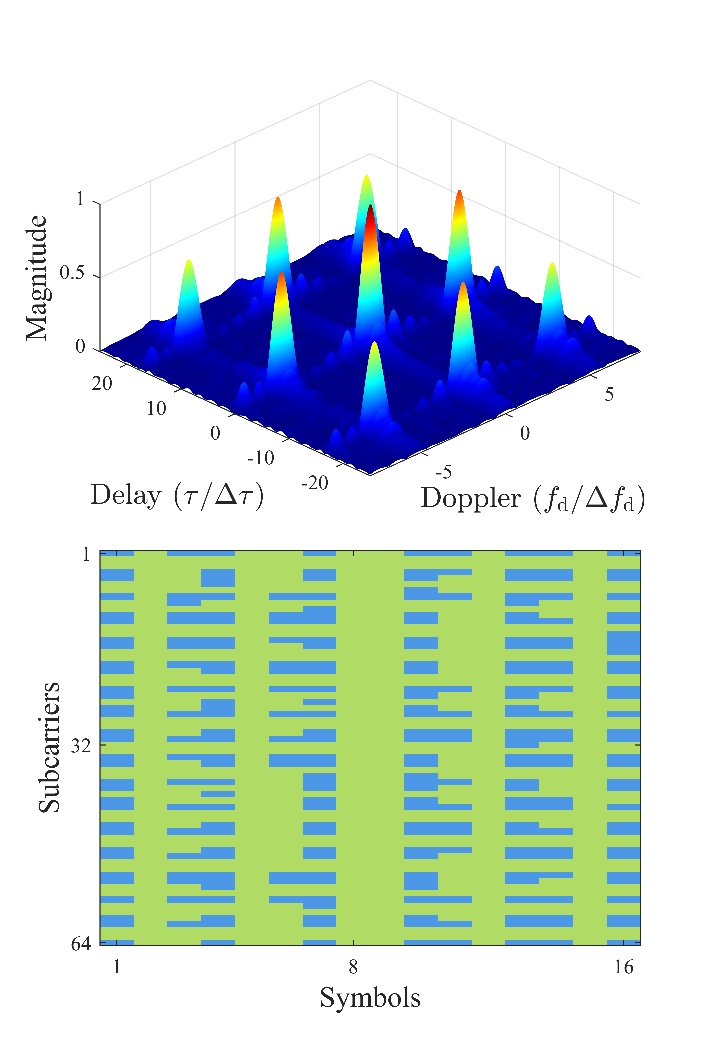} \label{fig:AF_RE_region1}
    }
    \hspace{-4mm}
    \subfigure[$l_{\text{max}} = 24$, $\nu_{\text{max}} = 6$.]{
        \includegraphics[width=0.2\linewidth]{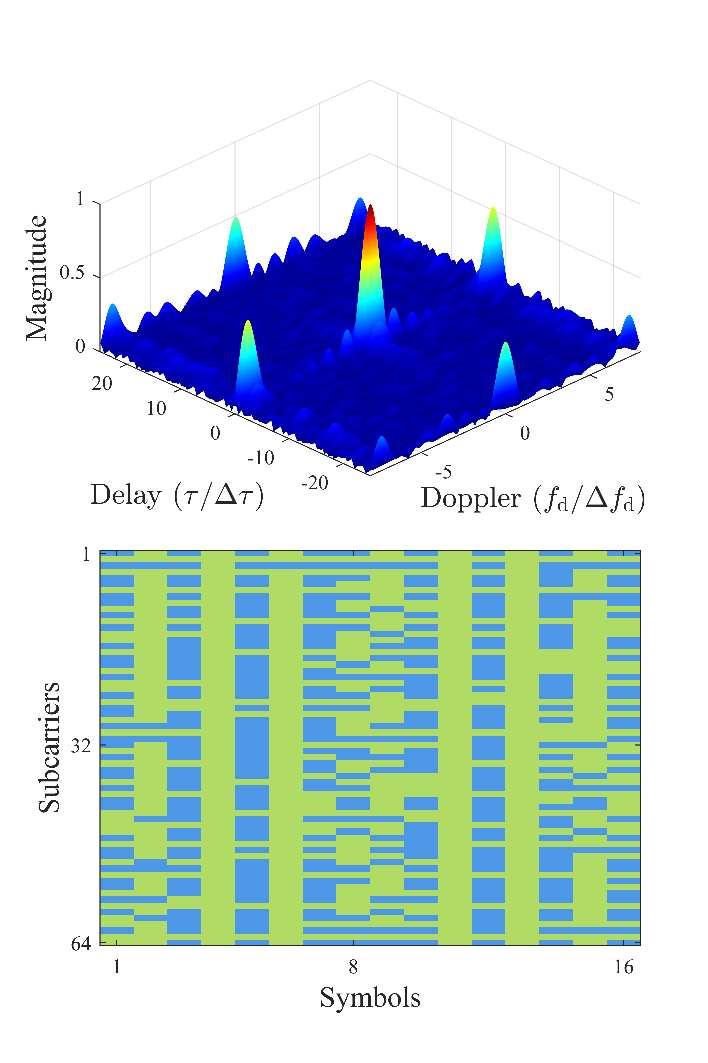} \label{fig:AF_RE_region2}
    }
    \hspace{-4mm}
    \subfigure[$-3$dB contour plots.]{
        \includegraphics[width=0.3\linewidth]{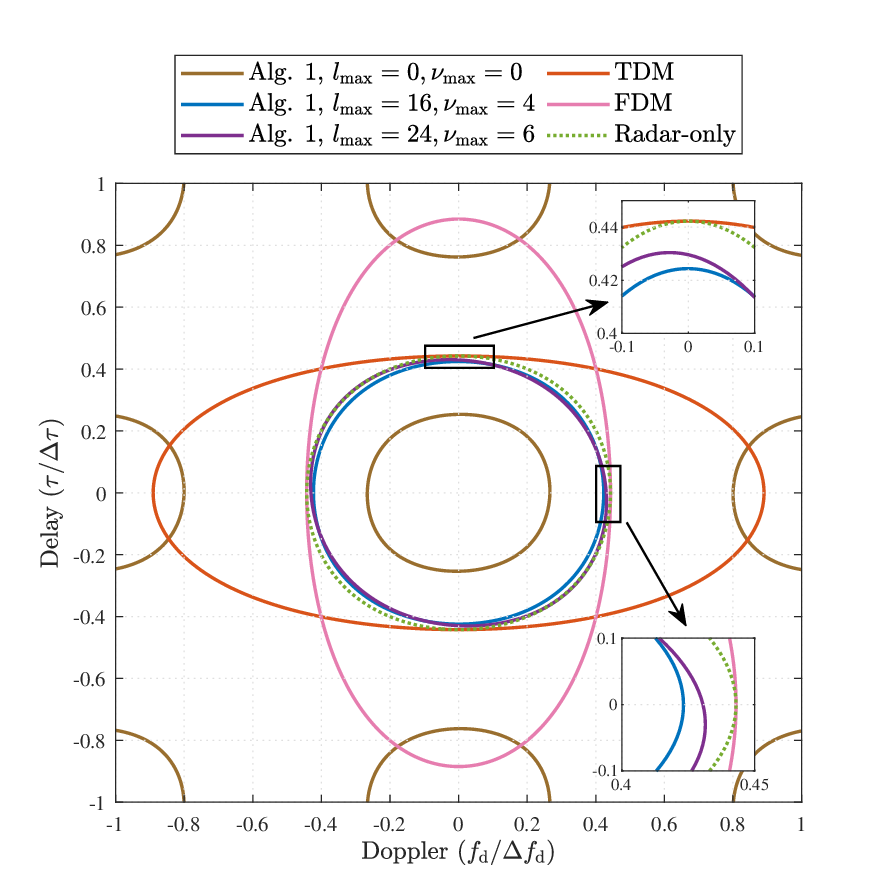} \label{fig:AF_3dB}
    }
    \centering
    \vspace{1mm}
    \caption{AFs and time-frequency resource allocation patterns for different sidelobe suppression regions (sensing: \textcolor{color_radar}{\rule{7pt}{7pt}}, communication: \textcolor{color_comm}{\rule{7pt}{7pt}}), along with the $-3$dB contour plots of the AFs.}\label{fig:AF_RE_alg1}
    \vspace{-2mm}
\end{figure*}

\begin{figure}[!t]
    \centering
    \includegraphics[width = 3.0 in]{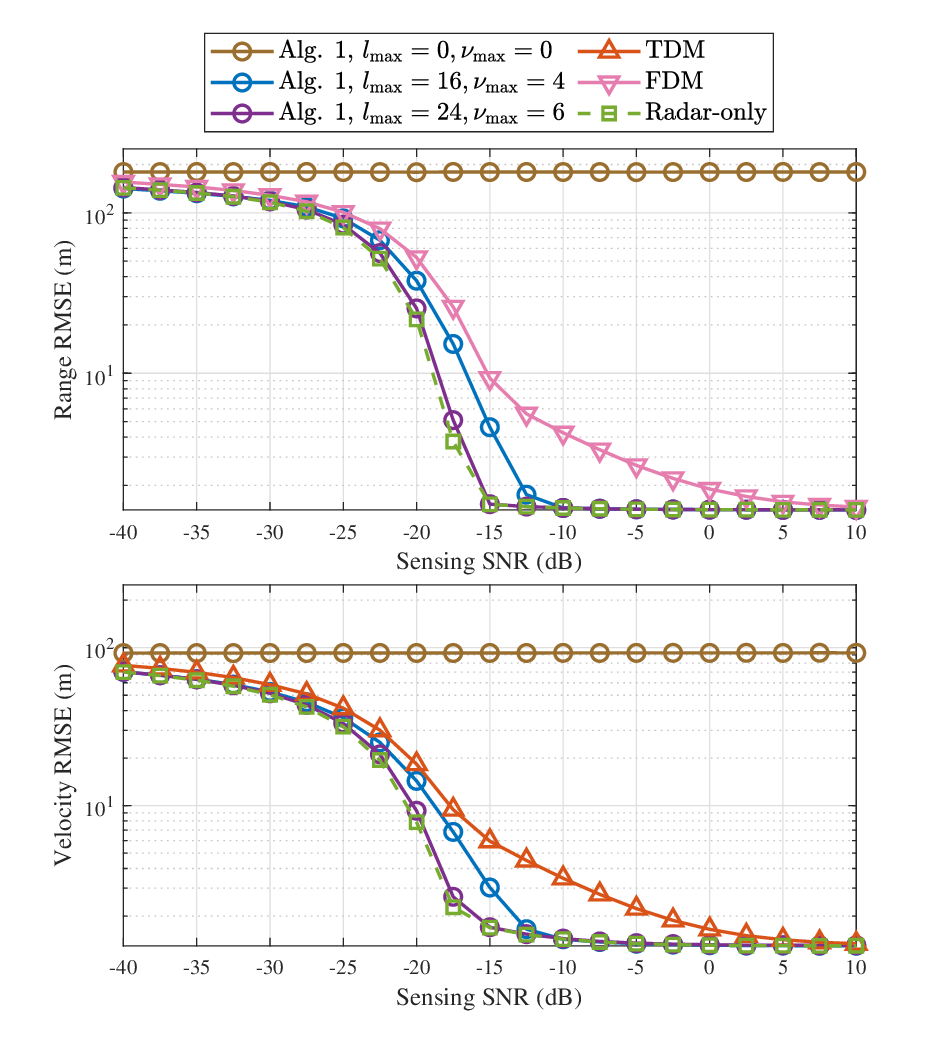}
    \vspace{-5mm}
    \caption{RMSE for range and velocity estimation versus sensing SNR, where the target RCS is set to $5$dBsm.}
    \label{fig:RMSE_Alg1}
    \vspace{-3mm}
\end{figure}

\subsection{Performance Evaluation of Resolution-Oriented Algorithm}
In this subsection, we evaluate the performance of the proposed resolution-oriented adaptive resource allocation design. We first illustrate the convergence performance of Algorithm \ref{Alg:minRes} in Fig. \ref{fig:converge_minRes}, where the objective value, range and velocity resolution versus the number of iterations under different settings are presented. We observe that Algorithm \ref{Alg:minRes} converges in a few iterations, with a monotonically decreasing sequence of objective and resolution values, which verifies the fast convergence and the effectiveness of the proposed algorithm.

To provide a qualitative assessment of the trade-offs between resolution, sidelobe suppression, and ambiguity in sensing performance, Fig. \ref{fig:AF_RE_alg1} illustrates the AF characteristics and resource allocation patterns for three different sidelobe suppression regions, where $l_{\text{max}}$ and $\nu_{\text{max}}$ represent the maximum delay and Doppler indices within the sidelobe suppression region. It is first observed from Fig. \ref{fig:AF_RE_woPSL} that, in the absence of the PSL constraint (i.e., $l_{\text{max}} = 0$, $\nu_{\text{max}} = 0$), the proposed scheme allocates only a limited number of REs at the edges of the time-frequency region, which results in severe ambiguities. In contrast, the PSL-constrained scheme effectively balances mainlobe sharpness and sidelobe suppression, as depicted in Figs. \ref{fig:AF_RE_region1} and \ref{fig:AF_RE_region2}, thereby enlarging the non-ambiguity region and improving sensing performance. Furthermore, as $l_{\text{max}}$ and $\nu_{\text{max}}$ increase, more time-frequency resources are required to effectively suppress the sidelobes.

To gain deeper insight into the resolution performance of the proposed algorithm, Fig. \ref{fig:AF_3dB} presents the $-3$dB contour plots of the AFs for various resource allocation approaches, including \textbf{TDM}, \textbf{FDM}, and \textbf{Radar-only}. The Radar-only approach allocates all REs exclusively to radar sensing. Compared to TDM and FDM, the proposed algorithm demonstrates superior performance in both delay and Doppler resolution. Notably, it slightly outperforms the radar-only allocation in terms of resolution but introduces ambiguity outside the sidelobe suppression region, while the radar-only approach avoids ambiguities across the entire delay-Doppler domain. As the sidelobe suppression region expands, the resolution performance of the proposed scheme gradually degrades, highlighting the fundamental trade-off between mainlobe sharpness (resolution) and sidelobe suppression (interference and ambiguity).
 
\begin{figure}[!t]
    \centering
    \includegraphics[width = 3.0 in]{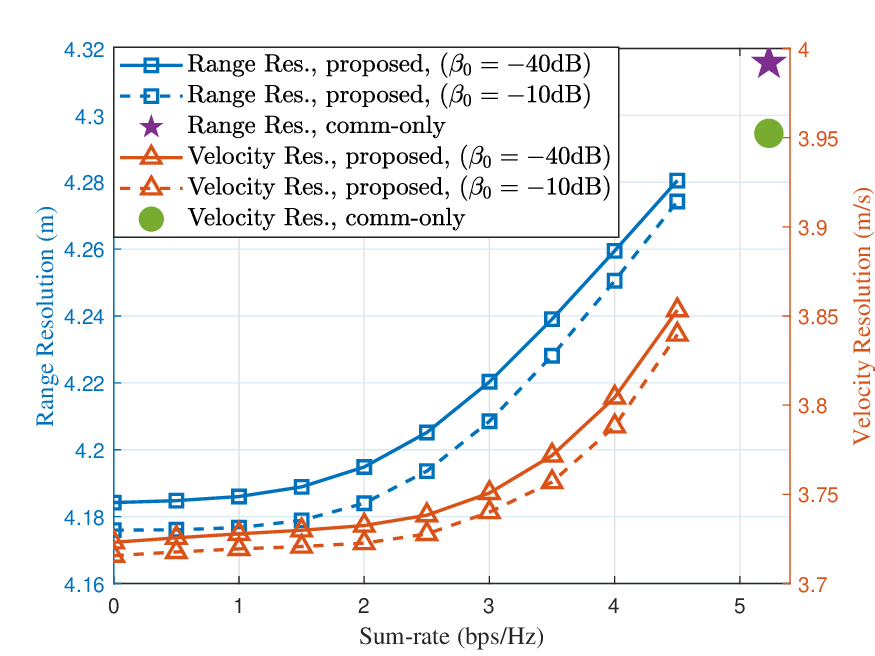}
    \caption{Range and velocity resolution versus communication sum-rate threshold $\eta_0$.}
    \label{fig:res_sumrate}
\end{figure}

Building on the above qualitative analysis, Fig. \ref{fig:RMSE_Alg1} quantitatively evaluates the parameter estimation performance by presenting the RMSE of the range and velocity estimates obtained using various resource allocation schemes as a function of sensing SNR. A single target is placed within the largest sidelobe suppression region ($l_{\text{max}} = 24, \nu_{\text{max}} = 6$). When the PSL constraint is not considered, the proposed approach exhibits significant range and velocity ambiguities, as shown in Fig. \ref{fig:AF_RE_woPSL}, leading to poor estimation performance. As the sidelobe suppression region expands, the range and velocity RMSE decreases rapidly, outperforming the TDM and FDM allocations. This performance improvement is attributed to the effective mitigation of parameter ambiguity, as demonstrated in Fig. \ref{fig:AF_RE_region1} and Fig. \ref{fig:AF_RE_region2}. Furthermore, under a sufficiently large sidelobe suppression region, the proposed adaptive design achieves nearly identical range and velocity RMSE as the radar-only approach.

\begin{figure}[!t]
    \subfigcapskip=-2pt
    \centering
    \hspace{-4mm}
    \subfigure[$\eta_0 = 1$bps/Hz.]{
        \includegraphics[width=0.52\linewidth]{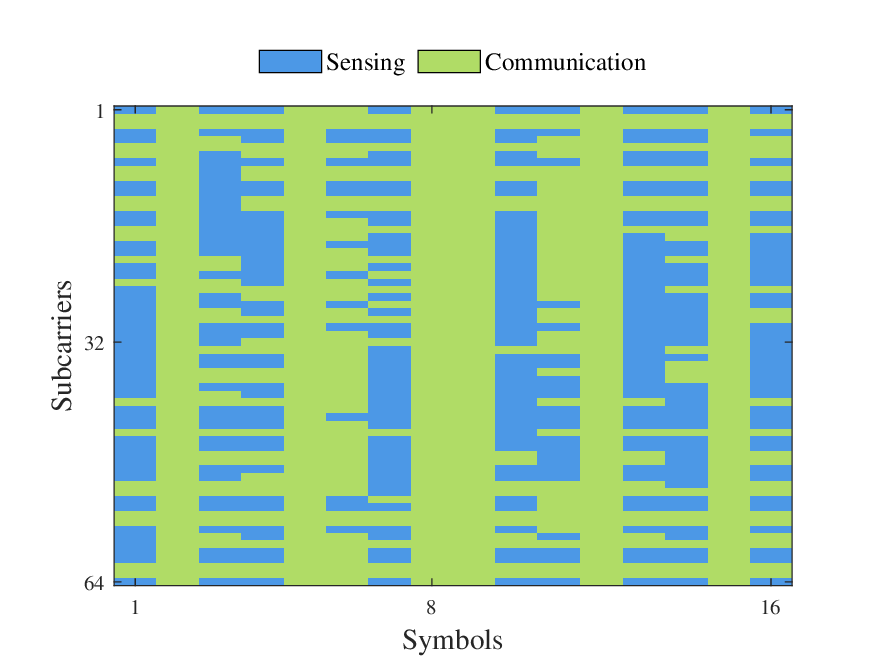}
    }
    \hspace{-8mm}
    \subfigure[$\eta_0 = 4.5$bps/Hz.]{
        \includegraphics[width=0.52\linewidth]{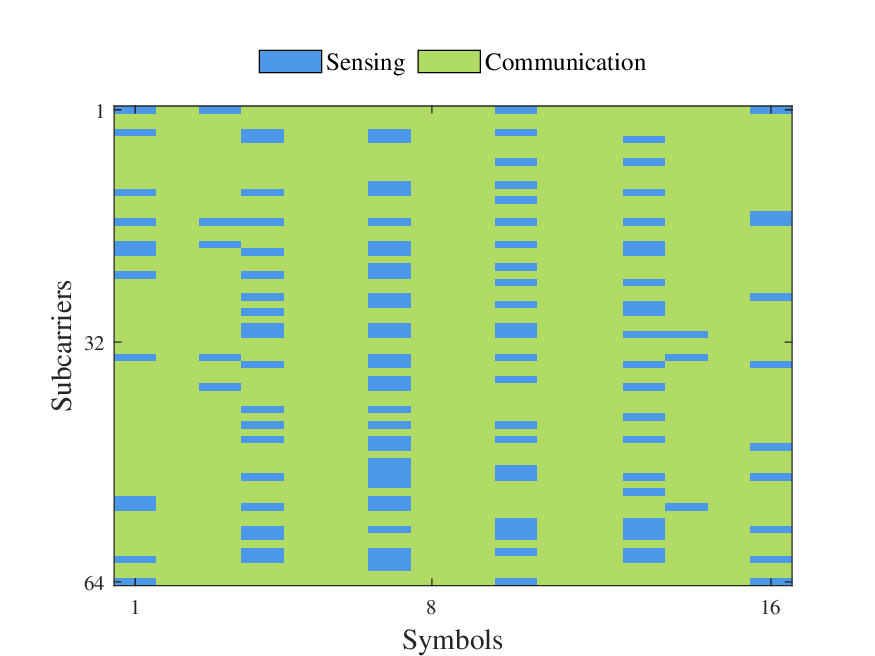}
    }
    \centering
    \vspace{-2mm}
    \caption{Resource allocation diagrams of the proposed Algorithm \ref{Alg:minRes} under different communication sum-rate thresholds $\eta_0$.}\label{fig:RE_alg1}
    \vspace{-2mm}
\end{figure}

Fig. \ref{fig:res_sumrate} illustrates the range and velocity resolution as a function of the communication sum-rate threshold $\eta_0$, where the \textbf{comm-only} approach refers to allocation of all REs for communications and employs a random $1024$-quadrature amplitude modulation (QAM) communication signal for sensing. As expected, both the range and velocity resolution degrade as the communication sum-rate threshold increases, highlighting the inherent trade-off between communication and sensing performance. Fig. \ref{fig:RE_alg1} depicts the resource allocation of Algorithm \ref{Alg:minRes} under different sum-rate thresholds $\eta_0$. With greater communication sum-rate demand (i.e., larger $\eta_0$), the sensing resource occupancy factor (RoF) decreases, while the communication RoF increases, further emphasizing the sensing-communication trade-off.

\begin{figure}[!t]
    \centering
    \includegraphics[width = 3.0 in]{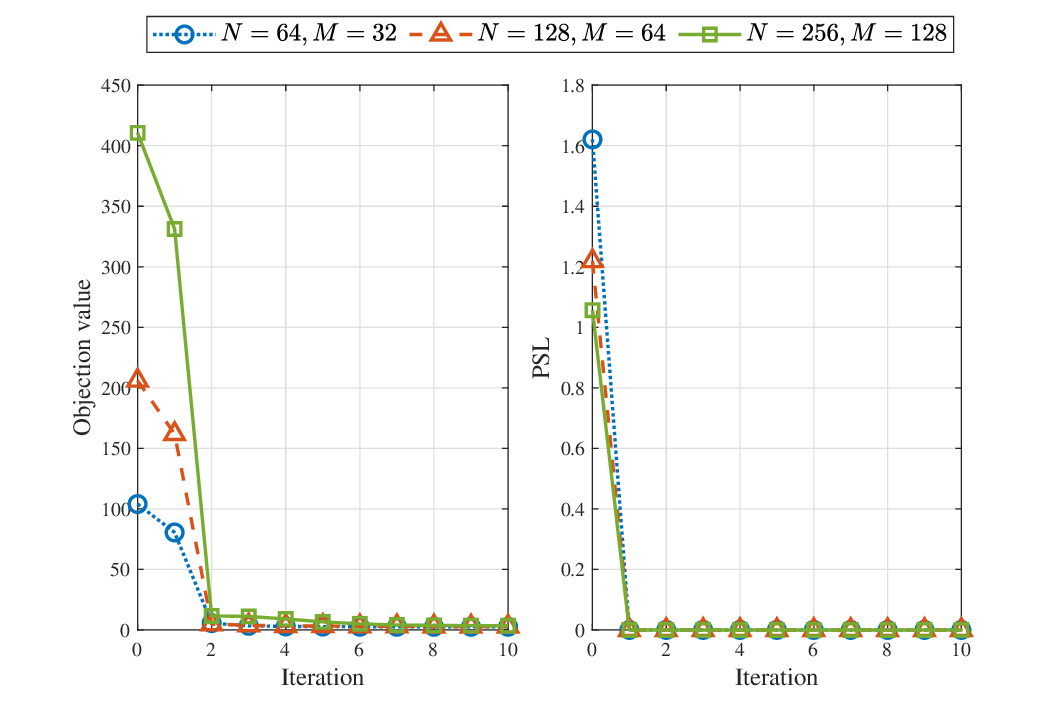}
    \caption{Convergence behavior of Algorithm \ref{Alg:minPSL}.}
    \label{fig:Alg2_convergence}
    \vspace{-3mm}
\end{figure}

\subsection{Performance Evaluation of Sidelobe-Oriented Algorithm}
In this subsection, we evaluate the performance of the proposed sidelobe-oriented adaptive resource allocation design. Fig. \ref{fig:Alg2_convergence} shows the convergence behavior of the proposed Algorithm \ref{Alg:minPSL} under different configurations. The objective value decreases rapidly and stabilizes within a small number of iterations, confirming the algorithm's efficiency. The monotonically decreasing objective and PSL values across all configurations validate the effectiveness of the proposed algorithm in achieving the desired performance.

\begin{figure}[!t]
    \subfigcapskip=-7pt
    \centering
    \subfigure[Alg. 2.]{
        \includegraphics[width=0.5\linewidth]{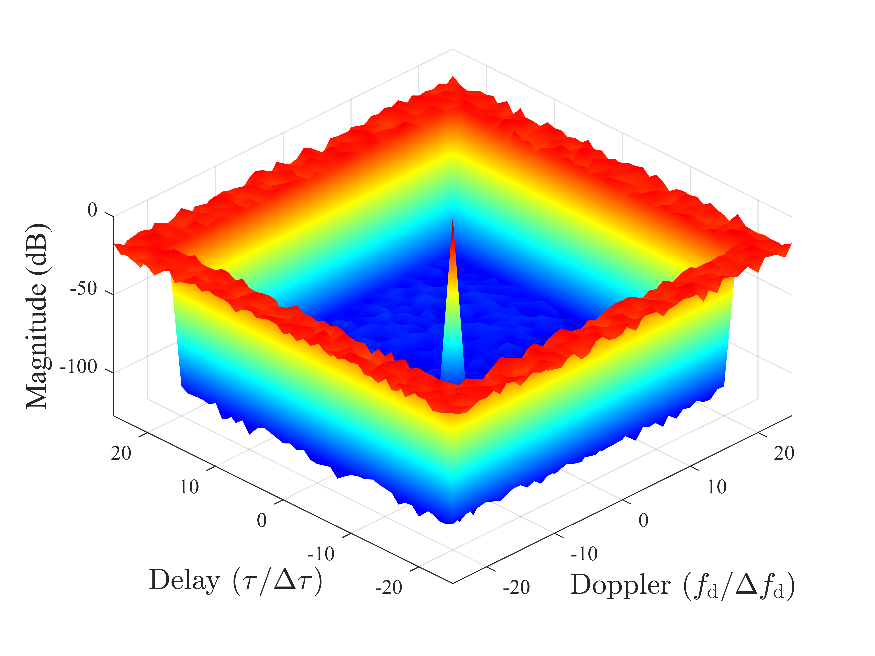} 
    }  
    \hspace{-8mm}
    \subfigure[Random.]{
        \includegraphics[width=0.5\linewidth]{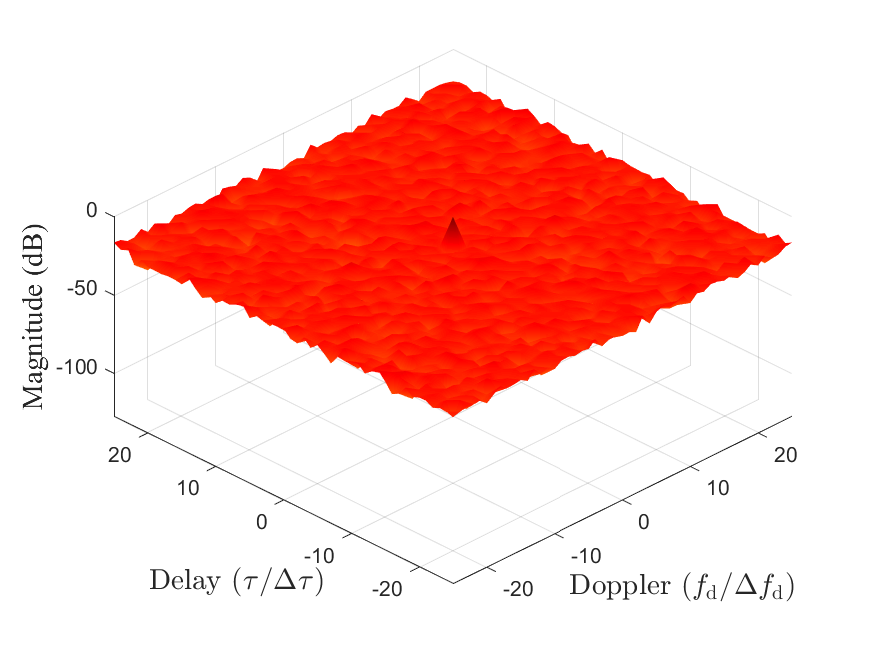} 
    }
    \centering
    \vspace{-2mm}
    \caption{AFs resulting from Algorithm \ref{Alg:minPSL} and random scheme.}\label{fig:AF_Alg2}
    \vspace{-6mm}
\end{figure}

\begin{figure}[!t]
    \centering
    \includegraphics[width = 3 in]{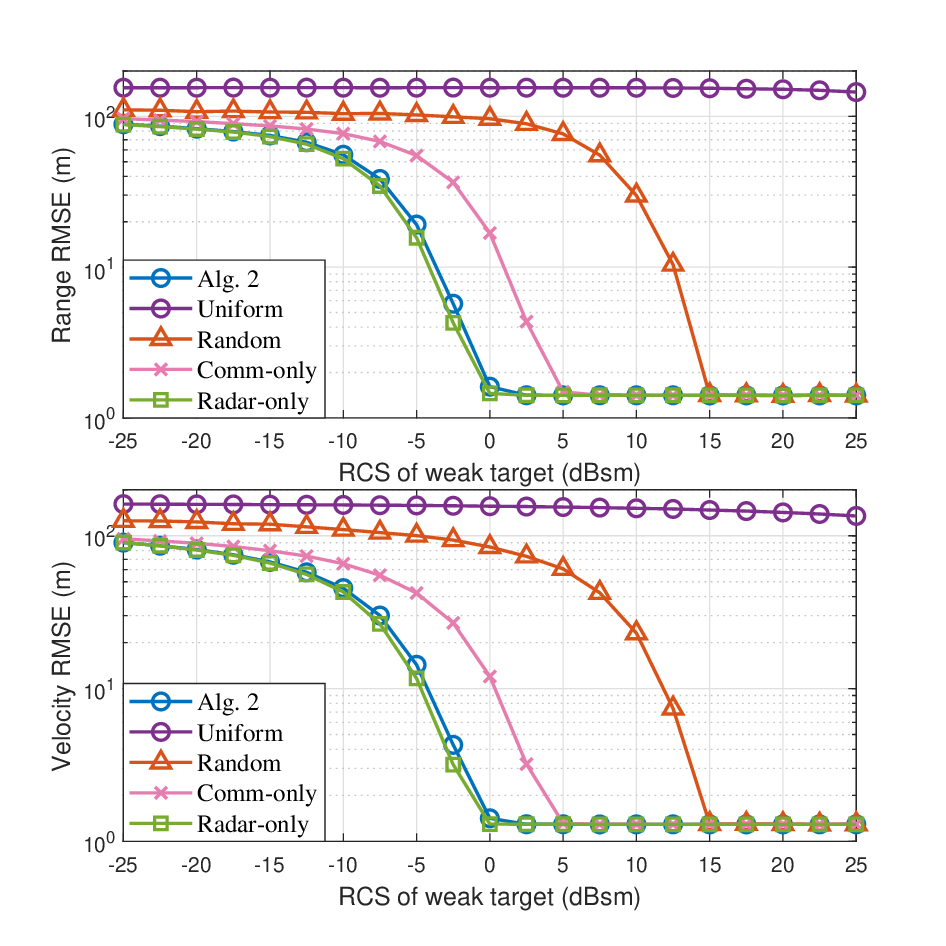}
    \vspace{-2mm}
    \caption{RMSE of range and velocity estimates versus the weak target RCS, where the RCS of the strong target is set to $5$dBsm.}
    \label{fig:RMSE_rcs_Alg2}
    \vspace{-3mm}
\end{figure}

Next, in Fig. \ref{fig:AF_Alg2} we evaluate the delay-Doppler sidelobe suppression performance by examining the AFs of the proposed Algorithm \ref{Alg:minPSL} and the random allocation. The proposed approach demonstrates superior sidelobe suppression in the delay-Doppler region of interest, with a reduction of over $90$dB compared to the random allocation, which randomly assigns the same number of sensing REs. This significant reduction mitigates mutual interference between closely spaced targets, thereby enhancing parameter estimation accuracy. Moreover, in contrast to the uniform allocation that exhibits severe range and velocity ambiguities as shown in Fig. 1(c), the proposed approach effectively eliminates these ambiguities within the desired delay-Doppler region.

The RMSE performance of the range and velocity estimates versus the RCS of the weak target is presented in Fig. \ref{fig:RMSE_rcs_Alg2}, where we compare the following approaches: \textbf{Uniform}, \textbf{Random}, \textbf{Comm-only}, and \textbf{Radar-only}. The proposed approach significantly outperforms the random and comm-only allocations, achieving the same RMSE at values of the target RCS that are $5$dBsm and $15$dBsm lower, respectively. In fact, the proposed algorithm achieves RMSE performance almost identical to the radar-only benchmark that ignores the communication sum-rate constraint.
This highlights the ability of the proposed approach to optimally mitigate sidelobe interference and resolve ambiguities. In contrast, the uniform allocation fails to reliably estimate the target parameters due to severe range and velocity ambiguities, while the random and comm-only schemes experience degraded RMSE performance, primarily due to high sidelobe levels. 

\begin{figure}[!t]
    \centering
    \includegraphics[width = 3 in]{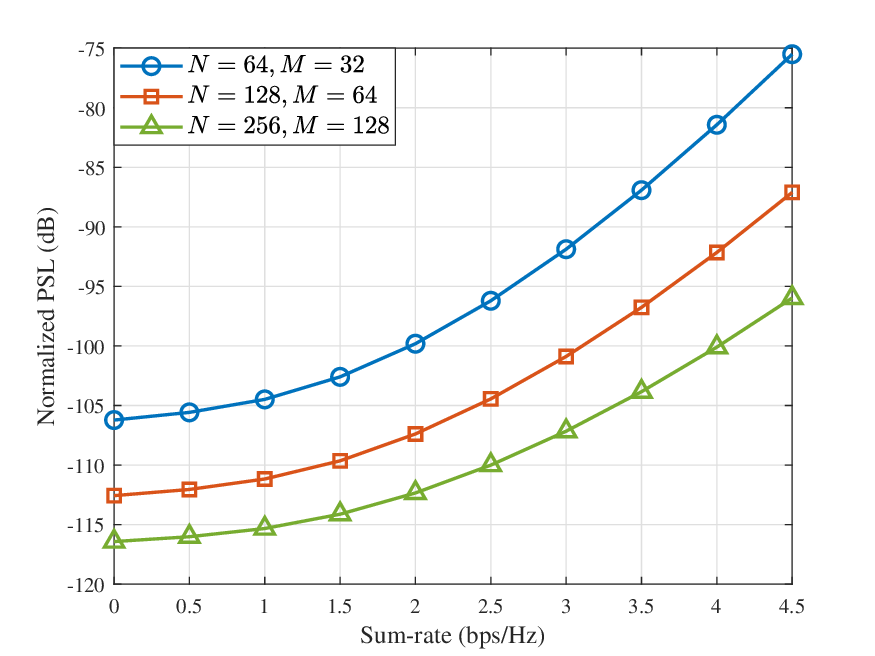}
    \vspace{-2mm}
    \caption{Normalized PSL versus the communication sum-rate threshold $\eta_0$.}
    \label{fig:PSL_sumrate}
    \vspace{-2 mm}
\end{figure}

\begin{figure}[!t]
    \subfigcapskip=-2pt
    \centering
    \hspace{-4mm}
    \subfigure[$\eta_0 = 1$bps/Hz.]{
        \includegraphics[width=0.52\linewidth]{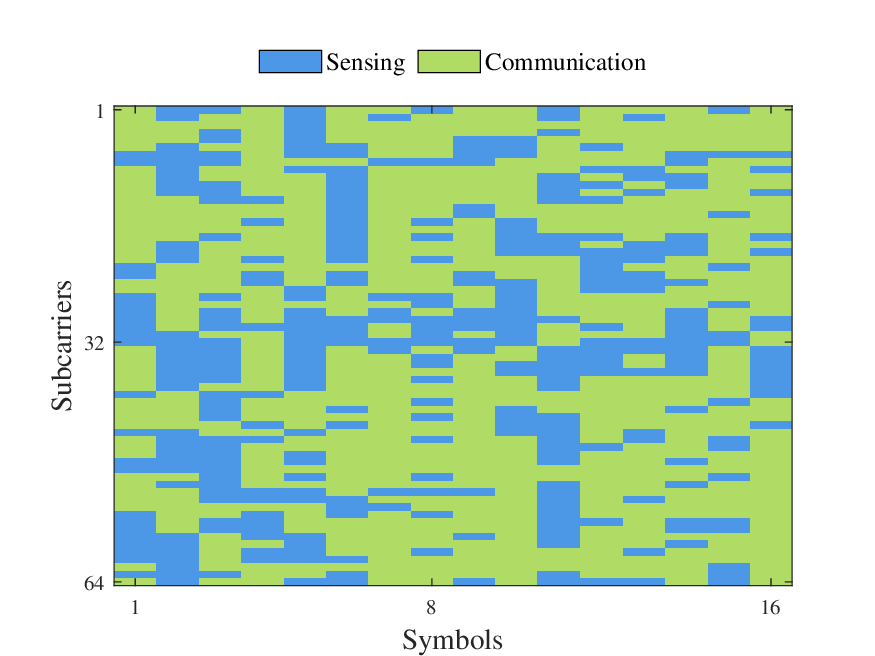}
    }
    \hspace{-8mm}
    \subfigure[$\eta_0 = 4.5$bps/Hz.]{
        \includegraphics[width=0.52\linewidth]{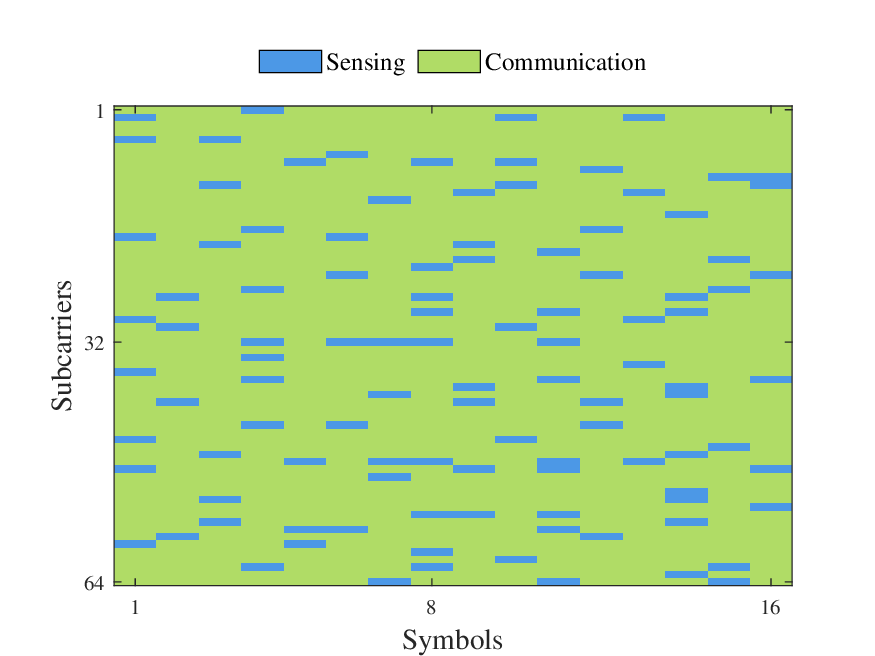}
    }
    \centering
    \vspace{-2mm}
    \caption{Resource allocation diagrams of the proposed Algorithm \ref{Alg:minPSL} under different communication sum-rate thresholds $\eta_0$.}\label{fig:RE_alg2}
    \vspace{-2mm}
\end{figure}

Fig. \ref{fig:PSL_sumrate} illustrates the inherent trade-off between sensing performance (normalized PSL) and communication performance (sum-rate) for the proposed approach. Given a total number of available REs, a higher communication sum-rate requirement results in degraded sensing PSL performance. Furthermore, increasing the number of available REs (i.e., larger $M$ and $N$) helps reduce the PSL while preserving spectral efficiency. To demonstrate the flexibility of the proposed adaptive allocation strategy, Fig. \ref{fig:RE_alg2} shows the resulting allocation patterns of Algorithm \ref{Alg:minPSL} under different communication sum-rate thresholds $\eta_0$. We see that, for larger $\eta_0$, more REs are allocated to communications, leading to a reduction in the sensing RoF. This highlights the algorithm's ability to adaptively allocate resources for various communication and sensing QoS requirements.

\section{Conclusion}
This paper has presented novel adaptive resource allocation designs for OFDM-ISAC systems. The proposed framework, along with the resolution-oriented and sidelobe-oriented algorithms, offers a robust approach for optimizing the power and subcarrier allocation in order to balance key sensing performance and communication QoS requirements. Extensive simulations validate the effectiveness of these strategies, demonstrating significant improvements in both sensing accuracy, resolution, and sidelobe suppression. These results highlight the potential of the proposed designs to effectively balance communication and sensing objectives, paving the way for more efficient and robust OFDM-ISAC implementations in next-generation wireless networks.

\appendices

\section{Proof of Proposition \ref{prop:AF_deri}}\label{appendix:AF_deri}
Substituting the OFDM radar transmit signal (\ref{eq:radar_tx_sig}) into (\ref{eq:AF_define}), the AF of the OFDM sensing signal	is given by
\begin{subequations} \label{eq:AF_deri}\begin{align}
& \chi(\tau,f_{\text{d}}) \notag \\
= & \frac{1}{MNT} \sum_{m_1, m_2} \sum_{n_1, n_2} u_{n_1, m_1}^0 u_{n_2, m_2}^0 s_{n_1, m_1}^{\text{r}} (s_{n_2, m_2}^{\text{r}})^{\ast} \notag                                     \\
& \times \sqrt{p_{n_1,m_1} p_{n_2,m_2}}   e^{-\jmath 2 \pi n_2 \Delta_f \tau}  \int_{-\infty}^{\infty} e^{\jmath 2 \pi (n_1-n2) \Delta_f t} \notag                                \\
& \times e^{\jmath 2 \pi f_{\text{d}}t} g\Big( \frac{t - m_1 T_{\text{sym}}}{T_{\text{sym}}} \Big)  g\Big( \frac{t + \tau - m_2 T_{\text{sym}}}{T_{\text{sym}}} \Big)   \text{d}t \\
= & \frac{1}{ MNT} \sum_{m} \sum_{n_1, n_2} u_{n_1, m}^0 u_{n_2, m}^0 s_{n_1, m}^{\text{r}} (s_{n_2, m}^{\text{r}})^{\ast} \sqrt{p_{n_1,m} p_{n_2,m}}  \notag     \\
& \times e^{\jmath 2 \pi n_2 \Delta_f \tau} \int_{t_{m,1}}^{t_{m,2}} e^{\jmath 2 \pi ((n_1\!-n_2) \Delta_f + f_{\text{d}}) t} \text{d}t, \label{eq:AF_cp_protection}
\end{align}\end{subequations}
where $t_{m,1} = mT_{\text{sym}}$, $t_{m,2} = (m+1)T_{\text{sym}} -T_{\text{CP}}$, and the inter-symbol interference for $m_1 \neq m_2$ is negligible in (\ref{eq:AF_cp_protection}) due to the protection provided by the CP. Since the Doppler shift is much smaller than the subcarrier spacing in practical applications, i.e., $f_{\text{d}} \ll \Delta_f$, we omit the cross-correlation term between different subcarriers $(n_1 \neq n_2)$ in (\ref{eq:AF_cp_protection}) due to the subcarrier orthogonality, i.e.,
\begin{equation}
\int_{t_{m,1}}^{t_{m,2}} \!e^{\jmath 2 \pi ((n_1-n2) \Delta_f + f_{\text{d}}) t} \text{d}t \approx \!\! \int_{t_{m,1}}^{t_{m,2}} \!e^{\jmath 2 \pi ((n_1-n2) \Delta_f) t} \text{d}t = 0.
\end{equation}
Therefore, we have
\begin{subequations}\begin{align}
& \chi(\tau,f_{\text{d}}) \notag \\
\approx & \frac{1}{ MNT}  \sum_{m = 0}^{M-1} \sum_{n = 0}^{N-1} u_{n, m}^0  p_{n,m} |s_{n,m}^{\text{r}}|^2 e^{-\jmath 2 \pi n \Delta_f \tau}\!\!\! \int_{t_{m,1}}^{t_{m,2}}\! e^{\jmath 2 \pi f_{\text{d}} t} \text{d}t  \label{eq:AF_sc_orthogonality}           \\
=  & \frac{\text{sinc}(f_{\text{d}}T)e^{\jmath \pi f_{\text{d}}(T\!-\! 2T_{\text{CP}})}}{ MN}\!\! \sum_{m = 0}^{M-1}\!  \sum_{n = 0}^{N-1}\! u_{n, m}^0  p_{n,m} e^{\jmath 2 \pi ( m f_{\text{d}} T_{\text{sym}}\!-\! n \Delta_f \tau)}   \label{eq:AF_zc_property} \\
\approx & \frac{1}{ MN} \sum_{m = 0}^{M-1} \sum_{n = 0}^{N-1} u_{n, m}^0  p_{n,m} e^{-\jmath 2 \pi n \Delta_f \tau} e^{ \jmath 2 \pi m f_{\text{d}} T_{\text{sym}}},
\end{align}\end{subequations}
where we leverage the constant modulus property of the Zadoff-Chu sequence, for which $|s_{n,m}|^2 = 1$, $\forall n,m$ in (\ref{eq:AF_zc_property}). Additionally, under the condition that $f_{\text{d}} T = f_{\text{d}} \Delta_f \ll 1$, we can approximate the value of $\text{sinc}(f_{\text{d}}T)e^{\jmath \pi f_{\text{d}}(T-2T_{\text{CP}})}$ as 1, resulting in a concise expression for the AF.

\vspace{2mm}
\section{Proof of Proposition \ref{prop:res_polyno}}\label{appendix:res_polyno}
Before converting the fractional function (\ref{eq:obj_simplify}) into a polynomial expression, we need to analyze the signs of $\mathbf{u}_0^T \mathbf{B} \mathbf{u}_0$ and $\mathbf{u}_0^T \mathbf{D} \mathbf{u}_0$ in (\ref{eq:obj_simplify}). Specifically,  the term $\mathbf{u}_0^T \mathbf{B} \mathbf{u}_0$ can be expanded as
\begin{subequations}
    \begin{align}
    & \mathfrak{I} \bigg\{ \sum_{m,m'}\sum_{n,n'} u_{n,m}^0 u_{n',m'}^0 p_{n,m} p_{n',m'} n' e^{\frac{\jmath \pi (n'-n)}{N}} \bigg\} \\
    = & \!\sum_{m,m'}\sum_{n,n'} \!u_{n,m}^0 u_{n',m'}^0 p_{n,m} p_{n',m'} n' \sin ((n'\!-\!n)\pi /N)  \\
    = & \!\sum_{m,m'} \! \sum_{n} \! \sum_{n'>n} \! u_{n,m}^0 u_{n',m'}^0 p_{n,m} p_{n',m'} n' \! \sin ((n' \!-\! n)\pi /N)  \notag    \\
    & \hspace{0.2cm} + u_{n',m'}^0 u_{n,m}^0 p_{n',m'} p_{n,m}  n \sin ((n-n')\pi /N)   \\
    = & \sum_{m,m'}\sum_{n} \sum_{n'>n}  u_{n,m}^0 u_{n',m'}^0 p_{n,m} p_{n',m'}  (n'-n)\notag  \\
    & \hspace{2cm} \times   \sin ((n'-n)\pi /N).
    \end{align}
\end{subequations}
For each pair $(n, n')$ with $n' > n$, both $(n' - n)$ and $\sin(\pi (n' - n)/N)$ are positive. Thus, for any $\mathbf{u}_0 \neq \mathbf{0}$ and $\mathbf{p} \neq \mathbf{0}$, the summation exclusively consists of positive terms, which implies $\mathbf{u}_0^T \mathbf{B} \mathbf{u}_0 > 0$.
Similarly, the term $\mathbf{u}_0^T \mathbf{D} \mathbf{u}_0$ can also be expanded as
\vspace{-1mm}
\begin{equation}
    \begin{aligned}
         & \mathbf{u}_0^T \mathbf{D} \mathbf{u}_0 = \sum_{n,n'}\sum_{m} \sum_{m'>m}  u_{n,m}^0u_{n',m'}^0 p_{n,m} p_{n',m'} \\
         & \hspace{2.7cm} \times  (m'-m) \sin ((m-m')\pi /M).
    \end{aligned}
\end{equation}
Obviously, for any $\mathbf{u}_0 \neq \mathbf{0}$ and $\mathbf{p} \neq \mathbf{0}$, the summation exclusively consists of negative terms, which implies $\mathbf{u}_0^T \mathbf{D} \mathbf{u}_0 < 0$. Then, using Dinkelbach's transform \cite{Dinkelbach_ManageSci_1967}, the fractional function (\ref{eq:obj_simplify}) can be converted into the following polynomial function:
\begin{equation}
    \begin{aligned}
    & \epsilon_\tau \big( 2| \mathbf{u}_0^T (\mathbf{I}_M \!\otimes\! \boldsymbol{\Phi}_{\tau_0}) \mathbf{p} |^2 \!-\! |\mathbf{u}_0^T \mathbf{p} |^2 \!+\! c_{\tau} t_\tau \mathbf{u}_0^T \mathbf{B}\mathbf{u}_0 \big) + \\
    & (1-\epsilon_\tau) \big( 2| \mathbf{u}_0^T (\boldsymbol{\Psi}_{f_0}^H \!\otimes\! \mathbf{I}_N)\mathbf{p} |^2 \!-\! |\mathbf{u}_0^T \mathbf{p} |^2 \!+\! c_v t_v \mathbf{u}_0^T \mathbf{D} \mathbf{u}_0 \big),
    \end{aligned}
\end{equation}
where $t_{\tau} \in \mathbb{R}$ and $t_v \in \mathbb{R}$ are auxiliary variables.


\vspace{-1mm}
\begin{thebibliography}{99}
    \vspace{-1mm}
    \bibitem{LiuFan_JSAC_2022} F. Liu, Y. Cui, C. Masouros, J. Xu, T. X. Han, Y. C. Eldar, and S. Buzzi, ``Integrated sensing and communications: Towards dual-functional wireless networks for 6G and beyond,'' \textit{IEEE J. Sel. Areas Commun.}, vol. 40, no. 6, pp. 1728-1767, Jun. 2022.

    \bibitem{Cui_Netw_2021} Y. Cui, F. Liu, X. Jing, and J. Mu, ``Integrating sensing and communications for ubiquitous IOT: Applications, trends, and challenges,'' \textit{IEEE Netw.}, vol. 35, no. 5, pp. 158-167, Sep. 2021.

    \bibitem{LiuRang WCM 2023} R. Liu, M. Li, H. Luo, Q. Liu, and A. L. Swindlehurst, ``Integrated sensing and communication with reconfigurable intelligent surfaces: Opportunities, applications, and future directions,'' \textit{IEEE Wireless Commun.}, vol. 30, no. 1, pp. 50-57, Feb. 2023.

    \bibitem{Zhang_JSTSP_2021} J. A. Zhang, F. Liu, C. Masouros, R. W. Heath, Z. Feng, L. Zheng, and A. Petropulu, ``An overview of signal processing techniques for joint communication and radar sensing,'' \textit{IEEE J. Sel. Topics Signal Process.}, vol. 15, no. 6, pp. 1295-1315, Nov. 2021.

    \bibitem{Zhang_CST_2022} J. A. Zhang, M. L. Rahman, K. Wu, X. Huang, Y. J. Guo, S. Chen, and J. Yuan, ``Enabling joint communication and radar sensing in mobile networks - A survey,'' \textit{IEEE Commun. Surveys Tuts.}, vol. 24, no. 1, pp. 306-345, First Quart. 2022.

    \bibitem{LiuRang JSTSP 2022} R. Liu, M. Li, Y. Liu, Q. Wu, and Q. Liu, ``Joint transmit waveform and passive beamforming design for RIS-aided DFRC systems,'' \textit{IEEE J. Sel. Topics Signal Process.}, vol. 16, no. 5, pp. 995-1010, Aug. 2022.

    
    \bibitem{Sturm_Proc_2011} C. Sturm and W. Wiesbeck, ``Waveform design and signal processing aspects for fusion of wireless communications and radar sensing,'' \textit{IEEE Proc.}, vol. 99, no. 7, pp. 1236-1259, Jul. 2011.

    \bibitem{Hakobyan_SPM_2019} G. Hakobyan and B. Yang, ``High-performance automotive radar: A review of signal processing algorithms and modulation schemes,'' \textit{IEEE Signal Process. Mag.}, vol. 36, no. 5, pp. 32-44, Sep. 2019.

    \bibitem{Keskin_TSP_2021} M. F. Keskin, V. Koivunen and H. Wymeersch, ``Limited feedforward waveform design for OFDM dual-functional radar-communications,'' \textit{IEEE Trans. Signal Process.}, vol. 69, pp. 2955-2970, Apr. 2021.

    \bibitem{Du_TSP_2024} Z. Du, F. Liu, Y. Xiong, T. X. Han, Y. C. Eldar, and S. Jin, ``Reshaping the ISAC tradeoff under OFDM signaling: A probabilistic constellation shaping approach,'' \textit{IEEE Trans. Signal Process.}, vol. 72, pp. 4782-4797, Sep. 2024.  

    \bibitem{Fan_Liu_2025}  F. Liu \textit{et al.}, ``Uncovering the iceberg in the sea: Fundamentals of pulse shaping and modulation design for random ISAC signals,'' Jan. 2025, [Online]. Available: https://arxiv.org/pdf/2501.01721.

    \bibitem{Peishi_TWC_2025} P. Li, M. Li, R. Liu, Q. Liu, and A. L. Swindlehurst, ``MIMO-OFDM ISAC waveform design for range-Doppler sidelobe suppression,'' \textit{IEEE Trans. Wireless Commun.}, vol. 24, no. 2, pp. 1001-1015, Feb. 2025.

    \bibitem{Peishi_TVT_2025} P. Li, M. Li, R. Liu, Q. Liu, and A. L. Swindlehurst, ``Low range-Doppler sidelobe ISAC waveform design: A low-complexity approach,'' submitted to \textit{IEEE Trans. Veh. Technol.}.


    \bibitem{Sun_RadConf_2014} H. Sun, F. Brigui, and M. Lesturgie, ``Analysis and comparison of MIMO radar waveforms,'' in \textit{Proc. IEEE Int. Radar Conf.}, Lille, France, Mar. 2014.

    \bibitem{Sturm_SN_2013} C. Sturm, Y. Sit, M. Braun, and T. Zwick, ``Spectrally interleaved multi-carrier signals for radar network applications and multi-input multi-output radar,'' \textit{IET Radar, Sonar Navigat.}, vol. 7, no. 3, pp. 261-269, Mar. 2013.
    
    \bibitem{Hakobyan_TAES_2020} G. Hakobyan, M. Ulrich, and B. Yang, ``OFDM-MIMO radar with optimized nonequidistant subcarrier interleaving,'' \textit{IEEE Trans. Aerosp. Electron. Syst.}, vol. 56, no. 1, pp. 572-584, Feb. 2020.

    \bibitem{Bica_ICASSP_2019} M. Bic\u{a} and V. Koivunen, ``Multicarrier radar-communications waveform design for RF convergence and coexistence,” in \textit{Proc. IEEE Int. Conf. Acoustics, Speech, and Signal Process. (ICASSP)}, Brighton, U.K., May 2019, pp. 7780-7784.

    \bibitem{Shi_SenJ_2019} C. Shi, F. Wang, S. Salous, and J. Zhou, ``Joint subcarrier assignment and power allocation strategy for integrated radar and communications system based on power minimization,'' \textit{IEEE Sensors J.}, vol. 19, no. 23, pp. 11167-11179, Dec. 2019.

    \bibitem{Shi_SysJ_2021} C. Shi, Y. Wang, F. Wang, S. Salous, and J. Zhou, ``Joint optimization scheme for subcarrier selection and power allocation in multicarrier dual-function radar-communication system,'' \textit{IEEE Syst. J.}, vol. 15, no. 1, pp. 947-958, Mar. 2021.

    \bibitem{Chen_CL_2023} Y. Chen, G. Liao, Y. Liu, H. Li, and X. Liu, ``Joint subcarrier and power allocation for integrated OFDM waveform in RadCom systems,'' \textit{IEEE Commun. Lett.}, vol. 27, no. 1, pp. 253-257, Jan. 2023.

    \bibitem{ZhangFan_JSAC_2023} F. Zhang, T. Mao, R. Liu, Z. Han, S. Chen, and Z. Wang, ``Cross-domain dual-functional OFDM waveform design for accurate sensing/positioning,'' \textit{IEEE J. Sel. Areas Commun.}, vol. 42, no. 9, pp. 2259-2274, Sep. 2024.


    \bibitem{Chu_TIT_1972} D. Chu, ``Polyphase codes with good periodic correlation properties (Corresp.),'' \textit{IEEE Trans. Inf. Theory}, vol. 18, no. 4, pp. 531-532, Jul. 1972.

    \bibitem{Swerling_TAES_1997} P. Swerling, ``Radar probability of detection for some additional fluctuating target cases,'' \textit{IEEE Trans. Aerosp. Electron. Syst.}, vol. 33, no. 2, pp. 698-709, Apr. 1997.

    \bibitem{Braun_thesis_2014} M. Braun, ``OFDM radar algorithms in mobile communication networks,'' Ph.D. dissertation, Dept. Inst. Commun. Eng., Karlsruhe Inst. Technol., Karlsruhe, Germany, 2014.

    \bibitem{woodward_book_1953} P. M. Woodward, \textit{Probability and Information Theory With Applications to Radar}. London, U.K.: Pergamon Press, 1953.

    \bibitem{Levanon_book_2004} N. Levanon and E. Mozeson, \textit{Radar Signals}. Hoboken, NJ, USA: Wiley, 2004.

    \bibitem{Richards_book_2005} M. Richards, \textit{Fundamentals of Radar Signal Processing}. New York, NY, USA: McGraw-Hill, 2005.

    \bibitem{Wang_arxiv_2024} X.-Y. Wang, S. Yang, K. Meng, H.-Y. Zhai, and C. Masouros, ``On the fundamental trade-offs of time-frequency resource distribution in OFDMA ISAC.'' July 2024, [Online]. Available: https://arxiv.org/abs/2407.12628.

    \bibitem{Song_TSP_2016} J. Song, P. Babu, and D. P. Palomar, ``Sequence design to minimize the weighted integrated and peak sidelobe levels,'' \textit{IEEE Trans. Signal Process.}, vol. 64, no. 8, pp. 2051-2064, Apr. 2016.

    \bibitem{Wang_TSP_2014} X. Wang, E. Aboutanios, M. Trinkle, and M. G. Amin, ``Reconfigurable adaptive array beamforming by antenna selection,'' \textit{IEEE Trans. Signal Process.}, vol. 62, no. 9, pp. 2385-2396, May 2014.

    \bibitem{Liu_TWC_2024} R. Liu, M. Li, Q. Liu, and A. Lee Swindlehurst, ``DOA estimation-oriented joint array partitioning and beamforming designs for ISAC systems,'' \textit{IEEE Trans. Wireless Commun.}, to appear.

    \bibitem{SunY_TSP_2017} Y. Sun, P. Babu, and D. P. Palomar, ``Majorization-minimization algorithms in signal processing, communications, and machine learning,'' \textit{IEEE Trans. Signal Process.}, vol. 65, no. 3, pp. 794-816, Feb. 2017.

    \bibitem{Dinkelbach_ManageSci_1967} W. Dinkelbach, ``On nonlinear fractional programming,'' \textit{Manage. Sci.}, vol. 133, no. 7, pp. 492-498, Mar. 1967.

    \bibitem{Boyd_book_2004} S. Boyd, S. P. Boyd, and L. Vandenberghe, \textit{Convex Optimization.} Cambridge, U.K.: Cambridge University Press, 2004.

\end{thebibliography}
\end{document}